\documentclass[aps,prd,a4paper,preprintnumbers,amsmath,amssymb,superscriptaddress,twocolumn,floatfix,showpacs]{revtex4}
\usepackage{amssymb}

\usepackage{float,graphicx}

\begin{document}

\title{Microscopic and Macroscopic Behaviors of Palatini Modified Gravity Theories}

\author{Baojiu Li}
\email[Email address: ]{b.li@damtp.cam.ac.uk} \affiliation{DAMTP,
Centre for Mathematical Sciences, University of Cambridge,
Wilberforce Road, Cambridge CB3 0WA, UK}
\author{David F. Mota}
\email[Email address: ]{d.mota@thphys.uni-heidelberg.de}
\affiliation{Institut f\"ur Theoretische Physik, Universit\"at
Heidelberg,  D-69120 Heidelberg, Germany}
\author{Douglas J. Shaw}
\email[Email address: ]{shaw@maths.qmul.ac.uk} \affiliation{School
of Mathematical Sciences, Queen Mary, University of London, London
E1 4NS, UK}

\date{\today}

\begin{abstract}
We show that, within modified gravity, the non-linear nature of
the field equations implies that the usual na\"{i}ve averaging
procedure (replacing the microscopic energy-momentum  by its
cosmological average) is invalid. We discuss then how the
averaging should be performed correctly and show that, as a
consequence, at classical level the physical masses and geodesics
of particles, cosmology and astrophysics in Palatini modified
gravity theories are all indistinguishable from the results of
general relativity plus a cosmological constant. Palatini gravity
is however a different theory from general relativity and predicts
different internal structures of particles from the latter. On the
other hand, and in contrast to classical particles, the
electromagnetic field permeates in the space, hence a different
averaging procedure should be applied here. We show that in
general Palatini gravity theories would then affect the
propagation of photons, thus changing the behaviour of a Universe
dominated by radiation. Finally, Palatini theories also predict
alterations to particle physics laws. For example, it can lead to
sensitive corrections to the hydrogen energy levels, the
measurements of which could be used to place very strong
constraints on the properties of viable Palatini gravity theories.
\end{abstract}


\maketitle

\section{Introduction}
\label{sect:Introduction}

Extensions of General Relativity (GR) have always received a great
deal of attention. Such theories are motivated by quantum gravity
models and by the wish to find phenomenological alternatives to
the standard paradigm of dark matter and dark energy
\cite{carroll, skordis1, skordis2, bean1, bean2, li0, li1, hu1,
li2, halle, koivisto1, koivisto2, carroll2, koivisto3, brook,
clifton, ber, cog, hall, sot, bo, pop, amen, capo, steen, erica}.

Alternatives to Einstein gravity generally result in non-linear
corrections to the field equations. The application of these
equations to macroscopic (\emph{e.g.}, cosmological) scales
involves an implicit coarse-graining over the microscopic
structure of matter particles. However, when there are extra
non-linear terms in the field equations, \emph{a priori}, the
validity of the usual coarse-graining procedure can no longer be
taken for granted \cite{flan}. Hence, as discussed in \cite{flan,
dirk}, it may  be important to take into account the microscopic
structure of matter when applying the field equations to
macroscopic scales.

Unfortunately, up until now that has not been the common practice
\cite{pal1, pal2, pal3, pal4, li3}. This is probably because, in
GR as in Newtonian gravity, the microscopic structure of matter is
not particularly important on macroscopic scales. It is standard
practice to replace the metric, $g_{ab}$, and the energy momentum
tensor, $T_{ab}$, with some averages of them that coarse-grain
over the microscopic structure of matter particles. This simple
procedure only works, however, because on microscopic scales the
equations of GR are approximately linear.

In this \emph{article} we show that such an approach \emph{cannot}
simply be applied to modified gravity theories without a detailed
analysis of the energy-momentum microstructure. Indeed, the
na\"{i}vely averaging over the microscopic structure will
generally lead one to make incorrect predictions, and inaccurate
conclusions as to the validity of the theory \cite{flan}. Indeed,
it's possible for a theory that deviates significantly from GR at
the level of the microscopic field equations to be
indistinguishable from GR when correctly coarse-grained over
macroscopic (\emph{e.g.}, cosmological) scales. We illustrate this
point for a class of modified gravity theories in which the Ricci
scalar $R$ in the Einstein-Hilbert action is replaced by some
function $f(R, R^{ab}R_{ab})$.

It is well known that whenever such a replacement is made, the
field equations for the action can be derived according to two
inequivalent variational approaches: metric and Palatini. In the
former case, $R_{ab}$ and $R$ are taken to be constructed from the
metric $\bar{g}_{ab}$ which governs the conservation of the energy
momentum tensor, and the field equations are found by minimizing
the action with respect to variations in $\bar{g}_{ab}$. In the
alternative, Palatini approach, $R = R_{ab}\bar{g}^{ab}$ where
$R_{ab}$ is a function of some connection field $\Gamma^{a}_{bc}$
that is, \emph{a priori}, treated as being independent of
$\bar{g}_{ab}$. The field equations are then found by minimizing
the action with respect to both $\Gamma^{a}_{bc}$ and
$\bar{g}_{ab}$. If $f(R, R^{ab}R_{ab}) = R-2\Lambda$
(\emph{i.e.}~GR with a cosmological constant) then the two
approaches result in the same field equations. Otherwise they are
generally different.

Before going into further details there is one point to be
noticed. It has been argued that Palatini approach, as outlined
above, swaps one theory for another when applied to $f(R)$ actions
\cite{nottoPal}. Whether or not this is indeed the case, the
Palatini $f(R)$ field equations are mathematically equivalent to a
$\omega = -\frac{3}{2}$ Brans-Dicke theory with a potential, and
so certainly do correspond to \emph{a} mathematically valid, and
widely studied, modified gravity theory, even if it is not
technically derivable from an $f(R)$ action.

This work is organized as follows. In
\S~\ref{Sect:PalatiniGravity} we briefly introduce the main
ingredients of Palatini modified gravity theories, derive the
gravitational field equations for a general $f(R, R^{ab}R_{ab})$
Lagrangian and discuss their behaviour in vacuum. In
\S~\ref{sect:Averaging} we explain why the popular na\"{\i}ve
averaging procedure (in which one simply replaces the quantities
in the field equations with some coarse-grained averages) fails in
some cases, and detail our new averaging method. We then
reconsider the particle kinematics, cosmology, astrophysics and
atomic physics thoroughly using the new approach and compare our
results with old results in the literature in the subsequent
sections. In \S~\ref{Sec:ClassPart} the motion of classical
particles (clumps of energy density in tiny patches in between
which there is vacuum) is considered, and we find that particles
in Palatini theories move in exactly the same ways as they do in
GR, their active gravitational, passive gravitational and inertial
masses are all equal, and most importantly the predictions on
cosmology and astrophysics are also the same as those of GR. \S~\ref{sect:EMfield} is devoted to an analysis of the behaviour
of electromagnetic field in Palatini theories: in contrast to
classical particles the electromagnetic field permeates in the
space and its averaging is a bit different. We find that in
general Palatini theories (albeit not ones where $f(R, R^{ab}R_{ab}) =f(R)$) the propagation of photons is altered as
compared with GR, and the universe dominated by radiation will
also behave differently. \S~\ref{Sect:atom} then considers the
atomic physics. We argue that for atomic physics calculations it
is more convenient to work in the Einstein frame metric and show that
the matter Lagrangian is modified at the field theoretic level. In
particular, the atomic energy levels now depend on the
modification very sensitively and experimental data puts
very strong constraints on any Palatini-type deviations from GR.  Although the analysis is performed in the Einstein frame, we show that the resulting experimental constraints are independent of one's frame choice.  We finally
summarize in \S~\ref{Sect:summary}.

\section{Palatini $f(R, R_{ab}R^{ab})$ theories}

\label{Sect:PalatiniGravity}

In this section we briefly summarize the main ingredients of
Palatini modified gravity theories.

In general, to modify gravity one could add functions of the
curvature invariants $R, R^{ab}R_{ab}, R^{abcd}R_{abcd}$ to the
standard Einstein-Hilbert action. In the Palatini variational
approach, the case of $R^{abcd}R_{abcd}$ has not yet been explored
up to date, so in this paper we shall focus on the special class
of theories where $f=f(R, R^{ab}R_{ab})$, the cosmology of which has recently been the subject of much
interest. We stress that, as
mentioned in \S~\ref{sect:Introduction}, in these theories the
Ricci tensor $R_{ab}$ is constructed from the connection
$\Gamma^{c}_{ab}$ which is generally \emph{not} the
Levi-Civit$\mathrm{\grave{a}}$ connection of the matter metric
$\bar{g}_{ab}$, which is instead denoted by
$\bar{\Gamma}^{c}_{ab}$. In what follows we shall use several
different notations and for clarity we define them here. We use
$g_{ab}$ to denote the metric whose Levi-Civit$\mathrm{\grave{a}}$
connection is $\Gamma^{c}_{ab}$ and as such we have $R_{ab} =
R_{ab}(\Gamma) = R_{ab}(g)$, the Ricci scalar calculated from this
metric is $R \equiv g^{ab}R_{ab}$; in a similar way for the matter
metric $\bar{g}_{ab}$ we have $\bar{R}_{ab} \equiv
\bar{R}_{ab}(\bar{\Gamma}) = R(\bar{g})$ and $\bar{R} =
\bar{g}^{ab}\bar{R}_{ab}$. Besides this, we also need the mixed
contractions $\mathcal{R} = \bar{g}^{ab}R_{ab}$, $\mathcal{R}^{ab}
= \bar{g}^{ad}\bar{g}^{bc}R_{cd}$, $\mathcal{R}^{b}{}_{a} =
\bar{g}^{bc}R_{ac}$. We further define the covariant derivatives
$\nabla_{c}$ and $\bar{\nabla}_{c}$ to be compatible with the
connections $\Gamma$ and $\bar{\Gamma}$ respectively, \emph{i.e.},
$\nabla_{c}g_{ab} = \bar{\nabla}_{c}\bar{g}_{ab} = 0$. To be
consistent with these conventions we shall rename the $f(R,
R^{ab}R_{ab})$ theories as $f(\mathcal{R},
\mathcal{R}^{ab}R_{ab})$ theories from now on, and these new
notations clearly show that the theories at hand are neither
metric ones nor pure affine ones. It will become clear below how
these different quantities relate with each other. Note also that
we assume $R_{ab}$ to be a symmetric tensor (if it contains
antisymmetric part then the field equation will be spoiled).

\subsection{The Action}

For Palatini $f(\mathcal{R}, \mathcal{R}^{ab}R_{ab})$ gravity we start from the
following action
\begin{eqnarray} \label{Eq:R1_action}
S_{f(\mathcal{R}, \mathcal{R}^{ab}R_{ab})} &=& \int
d^{4}x\sqrt{-\bar{g}}\frac{1}{2\kappa}f(\mathcal{R},\mathcal{R}^{ab}R_{ab})\nonumber\\
&& + S_{\mathrm{matter}}(\bar{g}_{\mu\nu}, \psi_{i}),
\end{eqnarray}
where $\kappa = 8\pi G$ with $G$ being the gravitational constant; $c= \hbar  = 1$.
$S_{\mathrm{matter}}$ is the matter action depending only on the
matter metric $\bar{g}_{ab}$ and specific matter species
$\psi_{i}$, and not on $\Gamma^{a}_{bc}$.   This means that the
energy momentum tensor, defined as:
\begin{eqnarray}
\mathcal{T}_{ab} &=& -\frac{2}{\sqrt{-\bar{g}}}\frac{\delta
S_{\mathrm{matter}}}{\delta\bar{g}^{ab}}, \label{EM:natural}
\end{eqnarray}
 is conserved with respect to $\bar{g}_{ab}$ and the particle geodesics
are determined by the metric $\bar{g}_{ab}$.  The conservation law is
\begin{eqnarray}
\bar{\nabla}^{a}\mathcal{T}_{ab} &=& 0,\nonumber
\end{eqnarray}
where $\bar{\nabla}_{a} g_{bc} = 0$ as defined above.

\subsection{Field Equations}

For the sake of convenience and clearness, we shall define $\Phi =
\mathcal{R}$, $\chi = \mathcal{R}^{ab}R_{ab}$ and $K^{a}_{\ b} =
\mathcal{R}^{a}_{\ b}$ for a general
$f(\mathcal{R},\mathcal{R}^{ab}R_{ab}) \equiv f(\Phi,\chi)$ theory. Minimizing the action with respect to the
variations in the connection, $\Gamma^{a}_{bc}$, gives
\begin{equation}
\sqrt{-g}g^{ab} = \sqrt{-\bar{g}}\bar{g}^{ac}\left( f_{\Phi}
\delta^{b}{}_c + 2f_{\chi} K^{b}{}_{c}\right), \label{Metric:Eq}
\end{equation}
where $f_{,\Phi}=\partial f(\Phi, \chi)/\partial\Phi$,
$f_{,\chi}=\partial f(\Phi, \chi)/\partial\chi$, $g_{ab}$ is the
metric whose Levi-Civit$\mathrm{\grave{a}}$ connection is
$\Gamma^{a}_{bc}$. In $f(\mathcal{R})$ theories where $f_{,\chi} =
0$, or in the cases when $K^{b}{}_{c} \propto \delta^{b}{}_c$, the
two metrics are related \emph{conformally}: $g_{ab} =
f_{,\Phi}\bar{g}_{ab}$. More generally, the relationship between
the two metrics is a \emph{disformal} one.

We define the matrix $\mathbf{K}$ by $\mathbf{K}^{a}{}_b =
K^{a}{}_b$ and minimize the action with respect to variations in
$\bar{g}_{ab}$ to find
\begin{equation}
f_{,\Phi}\mathbf{K} + 2f_{,\chi} \mathbf{K}^2 -\frac{1}{2}f \mathbf{I} = \kappa \mathbf{T}, \label{Matrix:Eq}
\end{equation}
where $\mathbf{T}^{a}{}_{b} = \bar{g}^{ac} \mathcal{T}_{bc}$ and
$\mathbf{I}$ is the $(4\times4)$ unit matrix. $\Phi$ and $\chi$
are then given by the following algebraic relations
\begin{equation}
\Phi = {\mathrm{tr}}\ \mathbf{K}, \qquad \chi = {\mathrm{tr}}\
\mathbf{K}^2
\end{equation}
The trace of Eq.~(\ref{Matrix:Eq}) reads
\begin{equation}
f_{,\Phi} \Phi + 2f_{,\chi}\chi - 2f = \kappa \mathcal{T},
\label{Trace:Eq}
\end{equation}
where $\mathcal{T} = \mathcal{T}^{a}_{\ a}$. Defining
\begin{eqnarray}
\mathcal{Q}^2 &=& \frac{\sqrt{-g}}{\sqrt{-\bar{g}}}\ =\
f_{,\Phi}^2 \left[ \mathrm{det}\left(\mathbf{I} +
\frac{2f_{,\chi}}{f_{,\Phi}} \mathbf{K}\right)
\right]^{1/2},\nonumber
\end{eqnarray}
and using Eq.~(\ref{Metric:Eq}) it is straightforward to check
that
\begin{eqnarray}
\mathcal{Q}^2 R^{a}{}_{b} &=& \mathcal{Q}^2
g^{ac}R_{cb}\nonumber\\
&=& \bar{g}^{ac} f_{,\Phi} R_{cb} +
2f_{,\chi}\bar{g}^{ac}\bar{g}^{de}R_{eb}R_{cd} \nonumber
\\ &=& f_{,\Phi} K^{a}{}_b + 2f_{,\chi} K^{a}{}_c K^{c}{}_b,\nonumber
\end{eqnarray}
and so Eq.~(\ref{Matrix:Eq}) is equivalent to
\begin{eqnarray}
G^{a}{}_b(g) &=& R^{a}{}_b(g) - \frac{1}{2}R(g)
\delta^{a}{}_b \label{EinsteinMod:Eq} \\ &=& \kappa \hat{T}^{\mu}{}_{\nu}
\equiv \frac{1}{\mathcal{Q}^2} \left(\kappa \bar{g}^{ac}\mathcal{T}_{cb}
- V(\Phi,\chi)\delta^{a}{}_b\right]  \nonumber
\end{eqnarray}
where the potential, $V(\Phi,\chi)$, is given by
\begin{equation}
V(\Phi,\chi) = \frac{f_{,\Phi}\Phi + 2f_{,\chi}\chi
- f(\Phi,\chi)}{2\kappa} \label{Potential:Def}
\end{equation}

When written in terms of $g_{ab}$, the Palatini field equations,
Eq.~(\ref{EinsteinMod:Eq}), are essentially those of General
Relativity but with a modified source term given in terms of the
natural energy momentum tensor, $\mathcal{T}^{a}{}_{b}$, together
with a potential term Eq.~(\ref{Potential:Def}). This
correspondence between Palatini theories and modified source
theories is well known.

We could also use Eqs.~(\ref{Metric:Eq}, \ref{Matrix:Eq},
\ref{Trace:Eq}) to rewrite the Palatini field equations entirely
in terms of what we will refer to as the natural, or matter,
metric $\bar{g}_{ab}$ (since it is the metric that naturally
appears in the matter action). In this case the field equations
for a general Palatini $f(\mathcal{R},\mathcal{R}^{ab}R_{ab})$
theory are fairly unwieldy and so we only present the form they
take in the special case when $f = f(\mathcal{R})$.   Defining $F
=f_{,\Phi}$ we have:
\begin{eqnarray} \label{Eq:R1_MFE_Physical}
\bar{G}_{ab} &=& \frac{1}{F}\kappa\mathcal{T}_{ab} -
\frac{1}{2}\bar{g}_{ab}\left(\mathcal{R} - \frac{f}{F}\right)
+ \frac{1}{F}(\bar{\nabla}_{a}\bar{\nabla}_{b} -
\bar{g}_{ab}\bar{\square})F\nonumber\\
&& - \frac{3}{2F^{2}}\left(\bar{\nabla}_{a}F\bar{\nabla}_{b}F
-\frac{1}{2}\bar{g}_{ab}\bar{\nabla}^{c}F\bar{\nabla}_{c}F\right)
\end{eqnarray}
where $\bar{G}_{ab} \equiv \bar{R}_{ab} -
\frac{1}{2}\bar{R}\bar{g}_{ab}$ is the Einstein tensor constructed
from $\bar{g}_{ab}$ and $\bar{\square} =
\bar{\nabla}^{a}\bar{\nabla}_{a}$. Since this equation involves
second order derivatives of $F$ (or equivalently of
$\mathcal{T}$), it is generally difficult to solve while it is
often easier to solve the equations in terms of $g_{ab}$.
Additionally, the curvature of $\bar{g}_{ab}$ is often much
larger, even over small scales, than one would have expected based
on the behaviour of the metric in GR. Consider for instance the
trace of Eq.~(\ref{Eq:R1_MFE_Physical}):
\begin{equation}
\bar{R}(\bar{g}) = - \frac{\kappa \mathcal{T}}{F} -
2\left(\mathcal{R} - \frac{f}{F}\right) + \frac{3\bar{\square}F}{F}
- \frac{3 (\bar{\nabla}F)^2}{2 F^2}. \label{Physical:trace}
\end{equation}
Based on GR one might expect that if $F = f_{,\Phi} \approx 1$,
$\bar{R} \sim \mathcal{O}(-\kappa \mathcal{T})$, however from
Eq.~(\ref{Physical:trace}) it is clear that unless $F$, which
depends algebraically on $\mathcal{T}$, is only varying very
slowly, one may actually have $\bar{R} \gg -\kappa \mathcal{T}$,
and so that gravity, as described by curvature of $\bar{g}_{ab}$,
is actually much stronger than one would naturally expect. The
immediate upshot of this is that it may not be appropriate to take
$\bar{g}_{ab} \approx \eta_{ab}$ over laboratory scales as one
might normally expect to be possible. We discuss the important
implications of this later in \S~\ref{sec:physical:metric} and
return to field equations in terms of the metric $\bar{g}_{ab}$ in
\S~\ref{sec:physicalmetric}.

\subsection{Behaviour in Vacuum}

\label{Sec:Theory:Vacuum}

The behaviour of these theories in a vacuum is, as we shall see
later, of great importance. In vacuum,
$\bar{g}^{ac}\mathcal{T}_{bc} = 0$, or possibly
$\bar{g}^{ac}\kappa\mathcal{T}_{bc} = -\lambda_0
\delta^{a}{}_{b}$, $\lambda_0 = {\rm const}$, in either case
Eq.~(\ref{Matrix:Eq}) gives $\mathbf{K} = \Phi_0  \mathbf{I} /4$,
and so $\Phi = \Phi_0$, $\chi = \Phi_0^2/4$ where $\Phi_0$ is
given by Eq.~(\ref{Trace:Eq})
\begin{eqnarray}
f_{,\Phi 0}\Phi_0 + \frac{1}{2} f_{,\chi 0}\Phi_0^2 &=&
-4\lambda_0 + 2f_{0}. \label{PhiVac}
\end{eqnarray}
where we defined $f_0 = f(\Phi_0,\Phi_0^2/4)$, $f_{,\Phi 0} =
f_{,\Phi}(\Phi_0,\Phi_0^2/4)$ and $f_{,\chi 0} =
f_{,\chi}(\Phi_0,\Phi_0^2/4)$. Denoting $\mathcal{Q}_0 =
\mathcal{Q}(\Phi_0,\Phi_0^2/4)$ and $V_0 = V(\Phi_0,\Phi_0^2/4)$,
we have
\begin{eqnarray}
\mathcal{Q}_0 &=& \frac{\left(2f_0 - 4\lambda_0\right)}{\Phi_0}, \\
\kappa V_0 &=& \frac{f_0 - 4\lambda_0}{2},
\end{eqnarray}
and so from Eq.~(\ref{EinsteinMod:Eq})
\begin{eqnarray}
\kappa \hat{T}^{a}{}_b\left(\bar{g}^{ac}\kappa\mathcal{T}_{bc} =
-\lambda_0 \delta^{a}{}_{b}\right) &=& - \frac{\Phi_0^2}{8(f_0 -
2\lambda_0)} \delta^{a}{}_b\nonumber\\
&\equiv& -\Lambda_{\rm eff}(\Phi_0) \delta^{a}{}_b\nonumber
\end{eqnarray}
It is very well known that the vacuum field equations of
$f(\mathcal{R},\mathcal{R}^{ab}R_{ab})$ Palatini theories are
equivalent to those of General Relativity with an effective
cosmological constant $\Lambda_{\rm eff}(\Phi_0)$.  One should
appreciate however that, depending on the specific form of $f$,
(1) there may be more than one value of $\Phi_0$ that satisfies
Eq.~(\ref{PhiVac}) and so the vacuum may not be unique in these
theories; (2) the value of $\Lambda_{\mathrm{eff}}$ could be
either positive or negative.

\section{Why Averaging Matters?}

\label{sect:Averaging}

It is important to stress that the equations we present in the
above section are all \emph{microscopic} field equations, which is
to say that they are only certainly valid when one has taken into
account all of the microscopic structures in the distribution of
energy and momentum described by $\mathcal{T}_{ab}$. \emph{A
priori} there is \emph{no} reason to expect these equations to
remain valid if $\mathcal{T}_{ab}$ and $\mathcal{T}$ are replaced
by some coarse-grained average. In particular for non-relativistic
baryonic matter, there is no reason to believe that we can course
grain over the peaks in density centered on each nuclei. In
chameleon scalar field theories \cite{cham1, cham2} where there is
a strong coupling between the scalar field and matter, for
example, the effective macroscopic field equations do not
necessarily look like the microscopic ones (see
Ref.~\cite{DEReview} \S~IV).

In standard General Relativity this problem does not emerge,
because $R$ depends linearly on $\rho$, so the average value of
$R$ could be calculated simply by replacing the microscopic value
of $\rho$ with its average. This averaging only works, however,
because of the linear dependence of $R$ on $\rho$.

In the Palatini modified gravity theories however, as we have
seen, $R_{ab}$ and $\mathcal{R}$ do \emph{not} generally depend
linearly on ${\mathcal T}$ (and hence $\rho$). Furthermore because
$f(\mathcal{R}, \mathcal{R}^{ab}R_{ab})$ is a nonlinear function
of $\mathcal{R}$ and $\mathcal{R}^{ab}R_{ab}$, there is no reason
to expect the averaged value of $f$ in a region of space to be
equal to the $f$ of the averaged values of $\mathcal{R}$ and
$\mathcal{R}^{ab}R_{ab}$; indeed this will generally \emph{not} be
the case. Even so, in almost all of the literature the microscopic
field equations are simply assumed to apply on macroscopical
scales. For instance, when the cosmology of these theories is
discussed the microscopic field equations are solved with
$\mathcal{T}_{ab}$ replaced by its cosmological average. There is
no reason, \emph{a priori}, to expect an analysis conducted along
these lines to be valid. In fact as we shall show below, when the
averaging is performed properly, the actual coarse-grained
behavior of the theory is very different from what one would find
if the microscopic field equations were na\"{\i}vely applied to
macroscopic scales.

Now let us discuss in more detail how this na\"{\i}ve averaging
could lead to incorrect results. Consider a body, or a region of
space, that, microscopically, contains $N$ non-relativistic
particles (\emph{e.g.}~nuclei) each with (microscopically) uniform
density $\rho_c$ and each occupying a volume (or with an average
volume) $V_{\rm p}$. The body as a whole is taken to have volume
$V_{\rm t ot}$. The space in between the particles is assumed to
be filled with some diffuse (non-relativistic) substance with
microscopic energy density $\rho_{s}$ (\emph{e.g.}~a diffuse
electron cloud or, if $\rho_s=0$, just empty space).

For clarity we take the $f(\mathcal{R})$ case as an example. By
Eq.~(\ref{Trace:Eq}), inside the particle we have $\Phi =
\Phi_{c}$ such that
\begin{eqnarray}
f_{,\Phi}(\Phi_{c})\Phi_{c} - 2f(\Phi_{c})\ =\ \kappa\mathcal{T}_{c}\
\approx\ -\kappa\rho_{c}\nonumber
\end{eqnarray}
where we have assumed $\rho_{c} \approx \mathrm{const.}$ and that
the particles are non-relativistic. Similarly in the space between
the particles we have instead $\Phi = \Phi_{s}$ where
\begin{eqnarray}
f_{,\Phi}(\Phi_{s})\Phi_{s} - 2f(\Phi_{s})\ =\ \kappa\mathcal{T}_{s}\ =\
-\kappa\rho_{s}.\nonumber
\end{eqnarray}
It is now straightforward to work out the true volume averaged
value of $\Phi, f(\Phi)$ and $f_{,\Phi}(\Phi)$. By volume averaging of a
quantity $Q(x)$ we mean
\begin{eqnarray}
\langle Q\rangle &=& \frac{\int_{V_{\mathrm{tot}}}d^{3}x
Q(x)}{V_{\mathrm{tot}}}.\nonumber
\end{eqnarray}
It is then clear that
\begin{eqnarray}
\langle\rho\rangle &=& \rho_{s}\left(1 -
\frac{NV_{p}}{V_{\mathrm{tot}}}\right) +
\rho_{c}\frac{NV_{p}}{V_{\mathrm{tot}}},
\end{eqnarray}
and similarly
\begin{eqnarray}
\langle\Phi\rangle &=& \Phi_{s} +
(\Phi_{c}-\Phi_{s})\frac{NV_{p}}{V_{\mathrm{tot}}},\\
\langle f(\Phi)\rangle &=& f(\Phi_{s}) +
\left[f(\Phi_{c})-f(\Phi_{s})\right]\frac{NV_{p}}{V_{\mathrm{tot}}},\\
\langle F(\Phi)\rangle &=& f_{,\Phi}(\Phi_{s}) +
\left[f_{,\Phi}(\Phi_{c})-f_{,\Phi}(\Phi_{s})\right]\frac{NV_{p}}{V_{\mathrm{tot}}},
\end{eqnarray}
and so on. In general the averaged value of the quantity $Q(x)$ is
\begin{eqnarray} \label{Eq:Averaging}
\langle Q\rangle &=& Q_{s} +
(Q_{c}-Q_{s})\frac{NV_{p}}{V_{\mathrm{tot}}}\nonumber\\
&=& Q_{s} + (Q_{c}-Q_{s})\frac{\langle\rho\rangle -
\rho_{s}}{\rho_{c}-\rho_{s}}.
\end{eqnarray}
Generally the microscopic field equations give $Q$ as a function
of $\mathcal{T}\approx-\rho$, \emph{i.e.}, $Q(x) = Q(\rho(x)) =
Q(\rho)$. If one simply replaced $\rho$ by $\langle\rho\rangle$
then one would think that $\langle Q\rangle = Q_{a} \equiv
Q(\langle\rho\rangle)$. It is obvious however that unless $Q(\rho)
= a_{1} + a_{2}\rho$ (where $a_{1, 2}$ do not dependent on the
$\mathcal{T}_{ab}$ components) $\langle Q\rangle \neq Q_{a}$.

Because in the Palatini $f(\mathcal{R})$ theories $f$ is usually a
nonlinear function of $\mathcal{R}$, we conclude that, except for
isolated values of $\mathcal{R}$ we have $\langle
f(\mathcal{R})\rangle \neq f(\langle\mathcal{R}\rangle)$. As such
the effective macroscopic (\emph{i.e.}, coarse-grained) behavior
of Palatini $f(\mathcal{R})$ theories will not, as is almost
always assumed, be well described by the microscopic equations
with $\mathcal{T}_{ab}$ replaced by
$\langle\mathcal{T}_{ab}\rangle$. A generalization of the above
argument to the Palatini $f(\mathcal{R}^{ab}R_{ab})$ case is
straightforward if more complicated, due to the similar nonlinear
and algebraic nature of the field equations. As a result the
predictions made by applying the microscopic field equations in
Palatini modified gravity theories to macroscopic (\emph{e.g.},
astrophysical and cosmological) settings cannot generally be
trusted.

In the following sections we shall study the behaviors of
classical particles and radiation fields in Palatini gravity
theories when the averaging strategy Eq.~(\ref{Eq:Averaging}) is
used, and compare with the results obtained by na\"{\i}ve
averaging.

\section{Motion of Classical Particles}

\label{Sec:ClassPart}

In this section we consider the motion of a number of (classical)
microscopic particles in Palatini modified gravity theories. By a
classical particle, we simply mean some localized distribution of
energy and momentum, \emph{i.e.}, for particle $I$,
$\mathcal{T}_{ab} \neq 0$ only inside some world tube
$\mathcal{W}_{(I)}$. We further require that outside the world
tubes of the particles, gravity is \emph{weak}, which means that
gravity is, to a good approximation, Newtonian (or indeed
post-Newtonian). For this to be the case, we must require that the
typical separations between the particles are large compared to
their sizes. We also assume that these separations are large
enough so that the particles are effectively collisionless.  In
between the particles we have vacuum, \emph{i.e.},
$\mathcal{T}_{ab} = 0$ and so particles only interact with each
other gravitationally. In $f(\mathcal{R})$ theories, we can relax
this requirement to $\mathcal{T} = 0$ and so allow for
electromagnetic forces between the particles, however for the
moment we shall not do this.

We begin by considering a very simple, and rather idealized
version of this set-up, where all the particles are assumed to be
spherically symmetric, and furthermore, gravity is assumed to be
weak not only outside of the particle world tubes but also inside
them. This analysis was first presented in \cite{letter} and is
repeated here to serve as an illustration of the more general
result we present later in this work.

\subsection{Spherically symmetric particles}

\label{Sec:ClassPart:Spherical}

\subsubsection{A single Particle}

\label{subsect:1_particle}

We begin by considering a single, isolated, static and spherically
symmetric particle centered at $r=0$ and has a radius $r_p$.
Outside the particle $\mathcal{T}^{a}{}_{b} = 0$ and without loss
of generality we work in coordinates where
\begin{eqnarray}
g_{ab}\mathrm{d} x^a \mathrm{d} x^b &=& \mathrm{d} s^2\ =\
-e^{A(r,t)}\mathrm{d} t^2 +e^{B(r,t)}\mathrm{d} r^2 + r^2
\mathrm{d} \Omega^2.\nonumber
\end{eqnarray}
Because the particle is static, so $\mathcal{T}_{0}^{i} = 0$, and
the spherical symmetry imposes the conditions
$\mathcal{T}^{\theta}{}_{\theta} = \mathcal{T}^{\phi}_{\phi} =
p_{I}-p_{A}/2$, say, and $\mathcal{T}^{r}_{\phi} =
\mathcal{T}^{r}_{\theta} = 0$.  We define $\mathcal{T}^{r}{}_r =
p_{A}+p_{I}$, $\mathcal{T}^{0}{}_0 = -\rho$. One could find the
metric $\bar{g}_{ab}$ both inside and outside such a particle by
directly solving Eq.~(\ref{Eq:R1_MFE_Physical}), however, it is
much simpler to find $g_{ab}$ via Eq.~(\ref{EinsteinMod:Eq}), and
then if necessary one can recover the results in terms of
$\bar{g}_{ab}$ by some transformations, as we shall show in
\S~\ref{sec:physicalmetric}. Note that Eq.~(\ref{EinsteinMod:Eq})
is nothing but the standard Einstein equation with a modified
energy momentum tensor:
\begin{eqnarray}
\hat{T}^{a}{}_b &=&
\frac{1}{\mathcal{Q}^2}\left[\mathcal{T}^{a}{}_b - V(\Phi,
\chi)\delta^{a}{}_b\right],
\end{eqnarray}
We can see from this expression that $\hat{T}_{0}^{i} = 0$ if and
only if $\mathcal{T}_{0}^{i} = 0$, and that all the symmetries of
$\mathcal{T}^{a}{}_b$ are preserved by $\hat{T}^{a}{}_b$. The
solution for the metric functions in this case is well known.
Indeed if one works in terms of $\hat{T}^{a}{}_b$ rather than
$\mathcal{T}^{a}{}_b$, the equations that must be solved are just
those of static, spherically General Relativity where outside the
particle,
$$\hat{T}^{a}{}_b = - \Lambda_{\rm eff}(\Phi_0) \delta^a{}_b.$$
Taking the following ansatz for the metric functions
$$e^{A} = W(r)e^{2C(r)}, \qquad e^{B} = 1/W(r),$$ then the $tt$
and $rr$ components of the Einstein equations are
\begin{eqnarray}
\frac{1}{r^{2}}\frac{\mathrm{d}}{\mathrm{d}r} \left[r(1-W)\right]
&=& \frac{1}{\mathcal{Q}^{2}}\kappa\left[\rho+V(\Phi,
\chi)\right],\nonumber\\
\frac{2}{r}\frac{\mathrm{d}C}{\mathrm{d}r} &=&
\frac{1}{\mathcal{Q}^{2}W}\kappa(\rho+p_{I}+p_{A}).\nonumber
\end{eqnarray}
The solutions are \cite{olmo}
\begin{eqnarray}\label{Eq:R1_spherical_solution1}
W(r) &=& 1 - \frac{2GM(r)}{r} - \frac{\Lambda_{\rm eff}(\Phi_0)}{3}r^2, \\
C(r) &=& \frac{\kappa}{2} \int^{r}_{0}\, \mathrm{d} r^{\prime} \,
r^{\prime}\frac{\rho + p_{I} + p_A}{\mathcal{Q}^2(\Phi,\chi) W(r^{\prime})} + C', \label{Ceqn0}\\
M(r) &=& 4\pi \int_{0}^{r}\,\mathrm{d} r^{\prime} r^{\prime\,2}
\frac{\rho + \Delta V(\Phi, \chi)}{\mathcal{Q}^2(\Phi,\chi)},
\label{Meqn}
\end{eqnarray}
where $C'$ is a constant of integration and
$$\Delta V = V(\Phi, \chi) - \mathcal{Q}_{0}^{2}V_{0}/\mathcal{Q}^2(\Phi,\chi).$$
Outside the particle $C$ is a constant given by $$C =
\frac{\kappa}{2} \int^{r_{p}}_{0}\, \mathrm{d} r^{\prime} \,
r^{\prime}\frac{\rho + p_{I} + p_A}{\mathcal{Q}^2(\Phi,\chi)
W(r^{\prime})} + C',$$ and for convenience we can set this
constant to be zero by redefining the time coordinate. Then $$C' =
- \frac{\kappa}{2} \int^{r_{p}}_{0}\, \mathrm{d} r^{\prime} \,
r^{\prime}\frac{\rho + p_{I} + p_A}{\mathcal{Q}^2(\Phi,\chi)
W(r^{\prime})}$$ and so finally $C$ can be written as
\begin{eqnarray}
C &=& -\frac{\kappa}{2} \int^{r_{p}}_{r}\, \mathrm{d} r^{\prime}
\, r^{\prime}\frac{\rho + p_{I} + p_A}{\mathcal{Q}^2(\Phi,\chi)
W(r^{\prime})}. \label{Ceqn}
\end{eqnarray}
Additionally, outside the particle $\rho = \Delta V = 0$ and so
$M(r) = M_{p} = M(r_p) = {\rm const}$.  The angular components of
Eq.~(\ref{EinsteinMod:Eq}) provide us the analogue of the
Tolman-Oppenheimer-Volkoff (TOV) equation:
\begin{equation}
r\frac{\mathrm{d} P_{\rm  eff}}{\mathrm{d} r} + \frac{\rho + p_{I}
+ p_{A}}{\mathcal{Q}^2 r W(r)} Y(r) =
-3\frac{p_{A}}{\mathcal{Q}^2},
\end{equation}
where $$P_{\rm eff} = (p_{I}-p_{A} - \Delta V)/\mathcal{Q}^2$$ and
$$Y(r) = (4\pi G P_{\rm eff}r^3 + GM - \Lambda_{\rm eff}r^3/3).$$
This last equation implies that $P_{\rm eff}$ is $\mathcal{C}_{0}$
continuous at $r=r_p$, and so $P_{\rm eff}(r_p) = 0$.  The key
thing to note is that, no matter what the internal structure of
the particle is, and no matter what the details of the Palatini
theory are, outside such a particle the metric is simply
Schwarzschild-de-Sitter, just as it is in GR with cosmological
constant $\Lambda_{\rm eff}$.  This is of course precisely what
one should expect given that we have assumed spherical symmetry
and the vacuum field equations are the same as GR with $\Lambda =
\Lambda_{\rm eff}$. As expected as it might be, this simple
observations carries with it an important corollary which should
be stressed: to the outside world the gravitational field of the
particle has precisely the same form as it does in GR.  The
gravitational mass, $M_p$, depends on the components of
$\mathcal{T}^a{}_b$ in a slightly more complicated manner than it
does in GR. However, these components \emph{cannot} be measured
gravitationally by an external observer, and only $M_p$ can be
measured. Indeed, as we shall show below, $M_p$ is, in all the
usual senses, the physical mass of the particle, \emph{i.e.}, it
is clearly the active gravitational mass but it is \emph{also}
equal to the passive gravitational and inertial masses of the
particle.

\subsubsection{Multiple Particles}

Above we reviewed the spacetime about a single spherically
symmetric particle in Palatini theories, and noted that to an
external observer it was, in all theories, simply
Schwarzschild-de-Sitter, just as in GR with a cosmological
constant. This would be true of any modified source gravity where
a vacuum form for the natural energy momentum tensor,
$\mathcal{T}^{a}_{\ b}$, corresponds to a vacuum form for the
modified energy momentum tensor, $\hat{T}^{a}{}_{b}$, \emph{i.e.},
\begin{eqnarray}
\mathcal{T}^{a}_{b} \propto \delta^{a}{}_{b} &\Leftrightarrow&
\hat{T}^{a}{}_{b} \propto  \hat \delta^{a}{}_{b}.\nonumber
\end{eqnarray}

We now make a fairly straightforward generalization to the case of
multiple particles. We work in the weak field limit and assume
that the relative motions of the particles are non-relativistic,
which means $\mathcal{P} \sim \mathcal{O}(\epsilon)$ where
$\partial^2 \mathcal{P}$ can be any one of $\kappa
\mathcal{T}^{0}{}_{0}/\mathcal{Q}^2,\,
\kappa\mathcal{T}^{i}{}_{j}/\mathcal{Q}^2,\, \kappa \Delta
V(\Phi,\chi)/\mathcal{Q}^2 \sim \mathcal{O}(\epsilon)$,
$\mathcal{P}^{i} \sim o(\epsilon)$ where $\partial^2
\mathcal{P}^{i} = \kappa\mathcal{T}^{i}{}_{0}/\mathcal{Q}^2 \sim
o(\epsilon)$ where $\epsilon$ is a small parameter.

Correspondingly we linearized the metric $g_{ab}$ as $$g_{ab} =
\eta_{ab} + h_{ab}$$ with $h_{00},\, h_{ij} \sim
\mathcal{O}(\epsilon)$ and $h_{0i} \sim \mathcal{O}(\epsilon)$.
The assumption that the system is non-relativistic is equivalent
to assuming that it is quasi-static, that is, for any quantity $A$
each time derivative is suppressed by a positive power of
$\epsilon$ relative to each spatial derivative.

The metric inside a particle (labelled $K$), which would, by
assumption, be spherically symmetric in the absence of any other
matter fields, centered at $x_{(K)}(t)$ can then be calculated
straightforwardly by linearizing the Einstein equation for
$g_{ab}$, Eq.~(\ref{EinsteinMod:Eq}).

We have chosen to linearize $g_{ab}$ rather than $\bar{g}_{ab}$,
because $g_{ab}$ essentially obeys the Einstein equation, though
with a modified source, and so provided this source is suitably
weak the deviations of $g_{ab}$ from $\eta_{ab}$ will also be
small; depending on the details of the theory the same may
\emph{not} be true of $\bar{g}_{\mu \nu}$ (see Sections
\ref{sec:physical:metric}, \ref{sec:physicalmetric}). We shall
neglect the effects of the effective cosmological constant,
$\Lambda_{\rm eff}(\Phi_0)$, since on scales that are very much
smaller than the cosmological horizon, it is known to have
negligible effect on particle motions.

To $\mathcal{O}(\epsilon)$ the ${}^{0}{}_0$ component of
Eq.~(\ref{EinsteinMod:Eq}) becomes:
$$
\frac{1}{2} h_{00,ii} = \frac{\kappa \left(\mathcal{T}^{0}{}_{0} -
\mathcal{T}^{i}{}_{i} + 2\Delta V\right)}{2\mathcal{Q}^2},
$$
where, inside the $K^{\rm th}$ particle we define
$\mathcal{T}^{0}{}_{0} = -\rho^{(K)}$ and to leading order
$$\mathcal{T}^{i}{}_{j} = \left(p_{I}^{(K)}-\frac{1}{2}p_{A}^{(K)}\right)
\delta^{i}{}_{j} + \frac{3}{2}p_{A}^{(K)}y^{i}_{(K)}y^{j}_{(K)}/
r_{(K)}^2.$$  Here $y^{i}_{(K)} = x^{i} - x_{(K)}^{i}(t)$ and
$r_{(K)} = \sqrt{y^{i}_{(K)}y^{i}_{(K)}}$ is the radial coordinate
centered at $x_{(K)}^{i}(t)$. Defining $2S = h_{00}$ we then have:
\begin{equation}
S_{,ii} = -\frac{4\pi G}{\mathcal{Q}^2}\left[\rho + 3p_{I}
- 2\Delta V(\Phi, \chi)\right]. \label{Seqn}
\end{equation}
Both inside the $K^{\rm th}$ and in the vacuum between the
particles, Eq.~(\ref{Seqn}) has the solution:
\begin{eqnarray}
S(x^{i},t) &&= \frac{GM_{(K)}(\vert x - x_{(K)}(t) \vert)}
{\vert x - x_{(K)} \vert} \nonumber \\ &&- C_{(K)}(\vert x
-x_{(K)}(t)\vert) + U_{(K)}(x^{i},t),\nonumber
\end{eqnarray}
where
\begin{equation}
U_{(K)}(x^{i},t) = \sum_{J \neq K} \frac{GM_{(J)}}{\vert x
- x_{(J)}(t)\vert},
\end{equation}
where $C_{(K)}(r_{(K)})$ and $M_{(K)}(r_{(K)})$ are given by
Eqs.~(\ref{Ceqn}) and (\ref{Meqn}) with $r \rightarrow r_{(K)}$,
$\rho  \rightarrow \rho^{(K)}$, $p_{I} \rightarrow p_{I}^{(K)}$
and $p_{A} \rightarrow  p_{A}^{(K)}$. $M_{(J)}$ is the total mass
of the $J^{\rm th}$ particle and is also given by Eq.~(\ref{Meqn})
with the appropriate substitutions in the limit $r \rightarrow
\infty$.

In a similar manner, from ${}^{i}{}_{j}$ component of
Eq.~(\ref{EinsteinMod:Eq}) we have, $h_{ij} = 2\Psi \delta_{ij}$
where to $\mathcal{O}(\epsilon)$ $\Psi_{,kk} = \kappa
\hat{T}^{0}{}_0/2$:
\begin{equation}
\Psi = U + \frac{GM(r_{(K)})}{r_{(K)}} - D(r_{(K)}),
\end{equation}
where for a particle with radius $r_{p}$
$$
D(r) = \frac{\kappa}{2} \int^{r_{p}}_{r} \mathrm{d} r^{\prime}\,
r^{\prime}\,\frac{\left(\rho + \Delta V\right)}{\mathcal{Q}^2},
$$
so outside the particles $D = 0$. Finally for $j_{0i}$ we fix the
gauge so that:
$$
h_{0j,j} = 4\Psi_{,0},
$$
and then the Einstein equations give
$$
h_{0i,kk} = -2\kappa \mathcal{T}^{i}{}_{0}
$$
We can now write down the metric outside all but the $K^{\rm th}$
particle:
\begin{eqnarray}
&&g_{ab}\mathrm{d} x^{a} \mathrm{d} x^{b}\nonumber\\
&=& -\left[1-\frac{2GM(r_{(K)})}{r_{(K)}} + 2C(r_{(K)}) -
2U(x^{i},t)\right]\mathrm{d}t^2 \\
&& + \left[1+2U_{(K)}(x^{i},t) + \frac{2GM(r_{(K)})}{r_{(K)}} -
2D(r_{(K)})\right]\mathrm{d} \textbf{x}^{2}\nonumber
\end{eqnarray}
Because outside the particles we have $C=D=0$, so the metric is
given simply as
\begin{equation}
g_{ab}\mathrm{d} x^{a} \mathrm{d} x^{b} =
-\left(1-2\Psi_{N}\right)\mathrm{d} t^2 + (1+2\Psi_{N})\mathrm{d}
\textbf{x}^{2}, \label{outmetric}
\end{equation}
where $$\Psi_{N} = U_{(K)} + GM_{(K)}/r_{(K)} =  \sum_{J}
GM_{(J)}/r_{(J)}.$$

\subsubsection{Motion of Particles}

It is clear from Eq.~(\ref{EinsteinMod:Eq}) that if we define
$$\kappa \tilde{T}^{a}{}_b = \kappa\hat{T}^{a}{}_b + \Lambda_{\rm eff} \delta^{a}{}_b$$ we have
\begin{equation}
\nabla_{a} \tilde{T}^{ab} = 0, \label{Econs}
\end{equation}
where $\nabla_{a} g_{bc} = 0$ and we used the fact that
$\nabla_{a}G^{ab}(g)=0$ with $G_{ab}(g)$ the Einstein tensor of
$g_{ab}$. After a lengthy manipulation of the Einstein equations
its is well known that one can extract the following relation
\cite{tolman}
\begin{equation}
\sqrt{-g} \nabla_{a} \tilde{T}^{ab} = \partial_{a}\left[
\mathbf{T}^{ab} + \mathbf{t}^{ab}\right] = 0, \label{Eq:Mass_ordinary_div}
\end{equation}
in which $\mathbf{T}^{ab} = \sqrt{-g} \tilde{T}^{ab}$,
$$\mathbf{t}^{a}_{\ b} = \frac{1}{2\kappa}\left[\mathbf{L}\delta^{a}_{\ b} +
2\sqrt{-g}\delta^{a}_{\ b}\Lambda_{\mathrm{eff}} -
\mathbf{g}^{cd}_{b}\frac{\partial\mathbf{L}}{\partial\mathbf{g}^{cd}_{a}}\right]$$
with
\begin{eqnarray}
\mathbf{g}^{ab}_{c} &\equiv&
\left(\sqrt{-g}g^{ab}\right)_{,c},\nonumber\\
\mathbf{L} &\equiv&
\sqrt{-g}g^{ab}\left(\Gamma^{c}_{ad}\Gamma^{d}_{bc} -
\Gamma^{c}_{ab}\Gamma^{d}_{cd}\right),\nonumber
\end{eqnarray}
and
\begin{eqnarray} \label{Eq:Mass_evaluation}
16\pi G\left[\mathbf{T}^{ab}+ \mathbf{t}^{ab}\right] &=&
\mathcal{H}^{acbd}_{,cd}
\end{eqnarray}
with
\begin{eqnarray}
\mathcal{H}^{acbd} &\equiv& \mathbf{g}^{ab}\mathbf{g}^{cd} -
\mathbf{g}^{bc}\mathbf{g}^{ad},\nonumber\\
\mathbf{g}^{ab} &\equiv& g^{ab}\sqrt{-g}.\nonumber
\end{eqnarray}
$\mathbf{t}^{a}{}_b$ is called the pseudo-tensor density of
gravitational energy and momentum and it is not a true tensor
density. In contrast $\mathbf{T}^a{}_b$ is a true tensor density.
Eq.~(\ref{Eq:Mass_ordinary_div}) is not a covariant equation
because $\mathbf{t}^{\nu}_{\ \mu}$ is not a tensor density and
that it involves the ordinary rather than tensorial divergence.
However, the ordinary divergence will be useful to obtain (for
finite systems) the analogues of conservation of energy and
momentum as in classical dynamics by integration over some large
enough space region.

Returning to Eq.~(\ref{Eq:Mass_ordinary_div}), suppose that the
$K^{\rm th}$ particle is in a finite region of space,
$\mathcal{V}_{K}$, (which is a reasonable assumption) that
contains only the $K^{\rm th}$ particle and which is enclosed by a
boundary, $\Sigma_{K}$, located in the surrounding empty space at
a sufficient distance so that in some open set of points including
the boundary gravity is weak. We can then choose coordinates
$(t,x^{i})$ such that $g_{ab} = \eta_{ab} + h_{ab}$ where $h_{ab}$
is small. With these coordinates we have from
Eq.~(\ref{Eq:Mass_ordinary_div}):
\begin{eqnarray}
\frac{\mathrm{d}}{\mathrm{d} t} \int_{\mathcal{V}_K} \mathrm{d}^3
\textbf{x}\, \left[\mathbf{T}^{0a} + \mathbf{t}^{0a}\right] = -
\int_{\mathcal{V}_K} \mathrm{d}^3 \textbf{x}\,\left[
\mathbf{T}^{ia} + \mathbf{t}^{ia}\right]_{,i}.
\end{eqnarray}
Defining $J^{a}_{(K)} = \int_{\mathcal{V}_K} \mathrm{d}^3
\textbf{x}\, \left[\mathbf{T}^{0a} + \mathbf{t}^{0a}\right]$ we
have
\begin{eqnarray}
\frac{\mathrm{d} J^{a}}{\mathrm{d} t} = -\oint_{\Sigma_{K}}
\mathrm{d}^2 \Sigma_{i} \,\mathbf{t}^{ia}. \label{evoEqn}
\end{eqnarray}
Using Eq.~(\ref{Eq:Mass_evaluation}) we see that $J^{a}_{(K)}$ is
itself given by a surface integral:
\begin{equation}
J^{a}_{(K)} = \frac{1}{2\kappa} \oint_{\Sigma_{K}} \mathrm{d}^2
\Sigma_{i}\, \mathcal{H}^{ac0i}_{,c}.
\end{equation}
Using Eq.~(\ref{outmetric}) we find that outside the particles to
$\mathcal{O}(\epsilon)$ (remember that $\mathbf{g}^{ab} =
g^{ab}\sqrt{-g}$)
$$
\mathbf{g}^{00} = -(1+4\Psi), \qquad \mathbf{g}^{ij} = \delta^{ij},
$$
and to leading order $\mathbf{g}^{0i} = -h_{0i}$ so
\begin{eqnarray}
J^{0}_{(K)} &=& -\frac{2}{\kappa} \oint_{\Sigma_{K}} \mathrm{d}^2
\Sigma_{i} \Psi_{,i} \nonumber \\ &=& M_{(K)} -\frac{2}{\kappa}
\oint_{\Sigma_{(K)}} U_{,i} \mathrm{d}^{2}\Sigma_{i}\nonumber\\
&=& M_{(K)}.\nonumber
\end{eqnarray}
Similarly we find after some manipulation
$$
J^{i}_{(K)} = - \int_{\mathcal{V}_{(K)}} \mathrm{d}^3 x\,
\partial_{t}\tilde{T}^{0}{}_{0} x^{i}
$$
and since $T^{0}_{0}$ depends on $x^{i}$ and $t$ only in the
combination $r = \vert x^{i}-x^{i}_{(K)}(t) \vert$ at leading
order we have
\begin{eqnarray}
J^{i}_{(K)} &=& \int_{\mathcal{V}_{(K)}} \mathrm{d}^3 x\,
\tilde{T}^0_{0,k} \dot{x}_{(K)}^{k} x^{i}  \nonumber \\ &=& -
\dot{x}_{0} \int_{\mathcal{V}_{(K)}} \mathrm{d}^3
\tilde{T}^{0}{}_{0} \nonumber \\ &=&  \dot{x}_{(K)}^{i} M_{(K)}
\nonumber
\end{eqnarray}
So defining $u^{\mu}_{(K)} = (1, \dot{x}^{i}{}_{(K)})$, which is
the 4-velocity of the center of mass of the particle we have:
$$
J^{\mu}_{(K)} = M_{(K)} u^{\mu}_{(K)}.
$$
Now by calculating $\mathbf{t}^{ia}$ on $\Sigma_{(K)}$ and using
Eq.~(\ref{evoEqn}) find the evolution of $J^{\mu}_{(K)}$. The $0$
component of Eq.~(\ref{evoEqn}) gives
$$
\dot{M}_{(K)} = 0,
$$
and by noting that in the vacuum regions
$$
\mathbf{t}^{ij} = \frac{2}{\kappa} \left( \Psi_{,i} \Psi_{j}
- \frac{1}{2}\delta_{ij} \Psi_{,k}\Psi_{,k}\right).
$$
the $i$ components gives
\begin{eqnarray}
\frac{\mathrm{d}}{\mathrm{d} t} J^{i}_{(K)} &=& M_{(K)}
\frac{\mathrm{d}^2 x_{(K)}^{i}(t)}{\mathrm{d} t^2} =
-\frac{2}{\kappa}\int_{\mathcal{V}_{(K)}}\,\mathrm{d}^3\,x \,
\Psi_{,kk} \Psi_{,i} \nonumber \\ &=&
-\int_{\mathcal{V}_{(K)}}\,\mathrm{d}^3\,x \, \hat{T}^{0}{}_{0} U_{,i} \\
&=&  \nonumber U_{,i}(x_{(K)}^{k}(t)) M_{(K)}.
\end{eqnarray}
where the last line follows from expanding $U_{,i}$ about $x^{i} =
x^{i}_{(K)}$, $U_{,kk} = 0$ inside the body, and
$\hat{T}^{0}{}_{0}$ being a function of $r_{(K)}$ only at leading
order. Thus to leading order in $\epsilon$ we have
\begin{equation}
\frac{\mathrm{d}^2 x^{i}_{(K)}(t)}{\mathrm{d} t^2} =
U_{,i}(x_{(K)}^{i}), \label{Floweqn}
\end{equation}
which is precisely the Newtonian equation of motion, and precisely
the same result that we find in General Relativity.  Since
$\bar{g}_{ab}$ is the metric that appears in the matter action, it
is clear that small particles will move along geodesics in
$\bar{g}_{ab}$. Now Eq.~(\ref{Floweqn}) tells us that small
particles will also move along geodesics in $g_{ab}$.  Since
inside the body, in general, $\bar{g}_{ab}$ and $g_{ab}$ are only
related disformally, and hence will have different geodesics, this
result may seem rather counter-intuitive. It can, however, be
understood in a fairly simple fashion.  Although the two metrics
are in general related disformally, we noted in Section
\ref{Sec:Theory:Vacuum}, that in \emph{vacuum} the two metrics are
related by a constant \emph{conformal} factor, \emph{i.e.}, up to
a rescaling of coordinates they are the same metric. Thus outside
the particles, the geodesics of the two metrics are the
\emph{same}. Inside a particle then the only differences between
the geodesics of the two metrics are due to essentially local
effects, \emph{i.e.}, they are due only to the local matter
content. Such local differences cannot lead to the particle
developing an overall acceleration, as this would constitute a
self-acceleration and hence a violation of energy and momentum
conservation. Both the natural, $\mathcal{T}^{a}{}_b$, and the
effective, $\hat{T}^{a}{}_b$, are conserved (with respect to
$\bar{g}_{ab}$ and $g_{ab}$ respectively). They also both vanish
outside of the particle implying that there is no flux of energy
or momentum in or out of the particle. Thus the total energy and
momentum inside the particle must be conserved, and there can be
no self-accelerations. It should therefore come as no surprise
that particles move along geodesics in $g_{ab}$.

\subsection{Generalization}

We have seen, as we first reported in \cite{letter}, that the
motion of spherically symmetric, classical particles in a general
Palatini $f(\mathcal{R}, \mathcal{R}^{ab}R_{ab})$ theory is
observationally indistinguishable from the motion of the same set
of particles in pure General Relativity with a cosmological
constant, $\Lambda_{\rm eff}$.  The reason for this is remarkably
simple, albeit unappreciated up to this point, and follows from a
couple of very well known phenomenon, both illustrated above. The
first is that the field equations of all Palatini theories are
equivalent to those of GR with some modified source
$\mathcal{T}^{a}{}_{b}$, when this is just an effective
cosmological constant. The second, is that, quite remarkably, GR
is in many ways holographic and specifically the equations of
motions for a localised distribution of energy and momentum
surrounded by vacuum can be derived by considering surface, rather
than volume, integrals over curvature components \cite{surfint,
flanmotion}. Indeed in General Relativity, even when a modified
source is present, the gravitational mass of an isolated particle,
as well as higher mass moments, can all be defined in terms of
surface integrals outside the body and hence identified with
parameters in the general vacuum solution \cite{flanmotion}.

In Ref.~\cite{flanmotion}, via a computational tour-de-force, the
Post-2-Newtonian equations of motion for a set of $N$ classical
particles (with arbitrary internal structure and no assumed
symmetry) where calculated using surface integrals.  The motion of
the particles was found to depend entirely on quantities defined
outside the particles which are naturally interpreted as the
different mass moments of the particles. Indeed the spherically
symmetric particle analysis given above follows from a spherical
case of the calculation done in Ref.~\cite{flanmotion}, and so the
results of Ref.~\cite{flanmotion} imply that all our conclusions
apply equally well to particles with no symmetry.

 On the largest scales, the higher mass moments play a relativity
insignificant r\^{o}le and the motion of a set of non-relativistic
classical particles is determined, to a good approximation,
entirely by their initial positions, velocities and their
gravitational masses.  As should come of no surprise then, one
does not need to know the internal structure of a particle to know
how it moves, one does not, for instance, need to know the precise
molecular or atomic structure of a clump of particle to predict
its motion under gravitational and other external forces.  This is
true for any theory that is equivalent with General Relativity
with a modified source provided that a vacuum form \emph{i.e.},
$\propto \delta^{a}{}_{b}$, in the natural energy momentum tensor,
$\mathcal{T}^{a}{}_{b}$, corresponds to a vacuum form in the
effective one, $\hat{T}^{a}{}_{b}$.  In all these theories then
the modification of the source term does not ultimately matter as
classical particles i.e. clumps of matter surrounded by vacuum,
still move as particles do in General Relativity. Particle motions
are, in so far as particle motions under gravity and other
external forces go, observationally indistinguishable.

\subsection{The Physical Mass of a Classical Particle}

\label{subsect:Masses}

In this section we consider the meaning of the physical mass of a
\emph{classical particle}.  In principle, there can be a number of
different quantities associated with a particle that will, in
different situations, play the r\^{o}le of its mass.  Firstly, we
have the \emph{active gravitational mass} which is a measure of
the strength of the gravitational field induced by a body.
Secondly, there is the \emph{passive gravitational mass}, which
determines the force that a body feels within a given
gravitational field.  Lastly, there is the \emph{inertial mass} of
a body, which determines how quickly a particle's momentum changes
when a force is applied.  In General Relativity the latter two are
manifestly equal. Furthermore, any theory in which the passive
gravitational and inertial masses are equal is said to satisfy the
\emph{weak equivalence principle}, and in the absence of any
non-gravitational external forces the trajectory of particles will
depend only on their position and velocity and not on their
composition. Additionally in the Newtonian limit of General
Relativity, the two types of gravitational mass are also equal
\cite{WillGravReview}, and any violations of this equality are
tightly constrained by experiments \cite{WillGravReview}.

We saw above the Newtonian limit of the equation of motion for a
classical, non-relativistic, particle in both General Relativity
and Palatini theories is the same.  Our analysis was for
spherically symmetric particles although the results of Ref.
\cite{flanmotion} readily extend this to all classical particles.

In general if one defines the center of mass, $x_{\rm cm}^{i}$ of
the classical particle, inside which gravity is weak, with
effective energy momentum tensor $\tilde{T}^{a}_{\ b}$ in the
standard way, we have:
\begin{equation}
x_{\rm cm}^{i} = \frac{M^{i}_{p}}{M_{p}},
\end{equation}
where
\begin{eqnarray}
M^{i}_p = -\int_{\mathcal{V}_{p}} \mathrm{d}^3 x^{\prime} \sqrt{-g}\, \tilde{T}^{0}{}_{0} x^{\prime\,i}, \\
M_{p} = -\int_{\mathcal{V}_{p}} \mathrm{d}^3 x^{\prime}
\sqrt{-g}\,\tilde{T}^{0}{}_{0}.
\end{eqnarray}
where $\mathcal{V}_{p}$ is the volume occupied by the particle. Then
$$J^{a} = M_{p} \frac{\mathrm{d} x^{a}_{\rm cm}}{\mathrm{d} t}, $$
where $x^{a}_{\rm cm} = (t, x^{i}_{\rm cm})$. In the absence of
non-gravitational external forces, \emph{i.e.}, $\nabla_{a}
\tilde{T}^{a}{}_{b} = 0$, we have $\dot{M}_p = 0$ and:
$$
\frac{\mathrm{d} J^{i}}{\mathrm{d} t}  = M_{p} \frac{\mathrm{d}^2
x^{i}_{\rm cm}}{\mathrm{d} t^2}= M_{p} U_{,i}
$$
where $U_{,i}$ is the Newtonian potential due to other particles:
\begin{equation}
U =\sum_{I} \frac{GM_{(I)}}{r_{(I)}},
\end{equation}
where $r_{(I)}$ is the distance from $x^{i}_{\rm cm}$ to the
center of mass of the $I^{\rm th}$, and $M_{(I)}$ is that
particle's mass.  It is clear from this relation that $M_{p}$ and
the $M_{(I)}$ are the active gravitational mass of their
respective particles.  It is also clear that
\begin{equation}
\frac{\mathrm{d}^2 x^{\rm cm}}{\mathrm{d} t^2 } = U_{,i}.
\label{weakequiv}
\end{equation}
Thus the trajectory of a classical particle is independent of its
mass and hence its composition, and hence two particles at the
same point would feel the same acceleration. If the difference in
the acceleration of two bodies at the same point vanishes in one
frame with one choice of coordinates then it obviously must vanish
in all frames. Furthermore it is independent of which of the
metrics, $g_{ab}$ or $\bar{g}_{ab}$, one is working with when
performing the calculation \footnotemark[1]. The absence of any
differential acceleration is precisely the statement of the weak
equivalence principle which in turn is equivalent to saying that
the passive gravitational mass and inertial mass of any body are
equal.

\footnotetext[1]{In fact one could apply
$\bar{\nabla}^{a}\mathcal{T}_{ab} = 0$ to a classical particle
which is in static configuration, and show that the TOV equation
is equivalent to the statement that the particle has no
self-acceleration. If one forgets the internal structure (pressure
gradients \emph{etc.}) of the particle, then the gradient of
$\Phi$ inside the particle would not be balanced and would appear
in the geodesic equation of the particle. As $\Phi$ depends on the
local energy density of the particle, this would lead to the
conclusion that the particle feels a self-force that depends on
its energy density or materials and so WEP is violated. However,
with TOV equation taken into account, one can find that the
pressure gradients and $\Phi$ gradients cancel, leaving no
self-force on the particle.}

We should not be surprised by this. The weak equivalence principle
certainly holds in GR, irrespective of the composition or internal
structure of the particles one considers, and so it must also hold
in  any theory which is equivalent to GR up to a modified source
and in which a vacuum form for the natural energy momentum tensor,
$\mathcal{T}^{a}{}_{b}$, corresponds to a vacuum form for the
modified source term, $\hat{T}^{a}{}_{b}$. Since, in so far as
particle motions go, the only difference between such a theory and
GR are in the internal composition of the particles.

If non-gravitational forces are present and associated with an
effective energy momentum tensor $T^{ab}_{f}$, then generally
$\nabla_{a} \tilde{T}^{ab} = -\nabla_{a} T^{ab}_{f} = f^{b}$.
Assuming that the gravitational field is still dominated by the
matter in the particles, we find by repeating the analysis of
Section \ref{Sec:ClassPart:Spherical}:
\begin{eqnarray}
\frac{\mathrm{d} J^{i}}{\mathrm{d} t} &=& M_{p} \frac{\mathrm{d}^2
x^{i}}{\mathrm{d} t^2} =  M_{p} U_{,i} + \int_{\mathcal{V}_{p}}
\sqrt{-g} f^{i} \label{forceeqn} \\ &=& M_{p} U_{,i} + F^{i}.
\nonumber
\end{eqnarray}
Thus $F^{i}$ is then identified as the total, non-gravitational,
external force on the system. It is clear from
Eq.~(\ref{forceeqn}) that $M_{p}$ is, in addition to being the
active gravitational mass, the inertial mass of the particle. Thus
we have that inertial and both types of gravitational mass are
equal to $M_{p}$ in Palatini theories. In all the usual senses,
\emph{i.e.}, gravitational and inertial, then $M_{p}$ is the
physical mass of the particle.  However $M_{p}$ is not, in
general, equal to the particle mass in the matter action,
$S_{m}(\Psi^{i}, \bar{g}_{ab})$.  If it where then $M_{p}$ would
depend only on $\mathcal{T}^{a}{}_{b}$, but $M_{p}$ is given by
Eq.~(\ref{Meqn}), and this manifestly depends not only on
$\mathcal{T}^{a}{}_{b}$ but also on $V(\Phi,\chi)$ and
$\mathcal{Q}(\Phi,\chi)$.

\subsection{The Physical Energy Momentum Tensor and Metric}

\label{sec:physical:metric}

In the study of Palatini theories, it is common practice to refer
to $\mathcal{T}^{a}{}_{b}$ as the `physical' energy momentum
tensor, and $\bar{g}_{ab}$ as the `physical' metric. This is
purely a convention inherited from GR: because if one selects a
local inertial frame defined at some point with respect to
$\bar{g}_{ab}$, then in a region around this point physics is
well-described by the special relativistic limit of the matter
action, \emph{i.e.}, $S_{m}(\Psi^{i},\eta_{ab})$. The length of
this region is essentially equivalent to the length scale over
which the spacetime appears to be flat.

In a general scalar tensor theory, it is usually the case that
although gravity is modified, its strength is still roughly the
same and hence the length scale of the curvature of spacetime is
no smaller than it usually is. Assuming that $\bar{g}_{ab} \approx
\eta_{ab}$ in a laboratory is therefore no more or less valid that
it is in General Relativity.  Under these circumstances attaching
the label `physical' to $\mathcal{T}^{\mu}{}_{\nu}$ and
$\bar{g}_{ab}$ seems reasonable. In Palatini theories (and as
$\omega \rightarrow -3/2$ in Brans-Dicke theories), however,
things are different. Here, if we treat $\bar{g}_{ab}$ as the
physical metric, then gravity is much stronger over very small
scales. Indeed one could view these theories as containing an
additional component to the gravitational force which is
infinitely strong but has zero range. The upshot of this is that
the curvature of $\bar{g}_{ab}$ is much larger than one might
naively expect, and so the length scale over which one can treat
$\bar{g}_{ab} \approx \eta_{ab}$ is much smaller. Indeed,
depending on the composition of the particles considered, it may
even be smaller than the spatial extent of the particles
themselves.

The presence of a new strong component to the gravitational force
means that even over laboratory scales we can not be sure that
physics will be well described by $S_{m}(\Psi^{i},\eta_{ab})$.
Attaching the label `physical' to $\mathcal{T}^{a}{}_{b}$ and
$\bar{g}_{ab}$ is therefore misleading. Indeed we saw above that
in the Newtonian limit, the physical mass of a particle was not
given by a volume integral over $\rho = -\mathcal{T}^{0}{}_{0}$
but by a volume integral over $-\tilde{T}^{a}{}_{b}$.

It is, arguably, more straightforward to treat $g_{ab}$ and
$\hat{T}^{a}{}_{b}$ as being `physical'. The gravitational side of
the theory is then simple General Relativity as evidenced by
Eq.~(\ref{EinsteinMod:Eq}).

In this case we could make the definition
$\tilde{S}_{m}(\tilde{\Psi}^{i}, g_{ab}) =
S_{m}(\Psi^{i},\bar{g}_{ab})$, where $\tilde{\Psi}^{i}$ are some
redefinitions of the original matter fields. Since now the (new)
matter Lagrangian depends on the metric $g_{ab}$ whose Einstein
tensor behaves exactly like in GR
[cf.~Eq.~(\ref{EinsteinMod:Eq})], and on laboratory scales we are
justified in taking $g_{ab} \approx \eta_{ab}$, so on these scales
special relativity applies as in GR and gravity plays negligible
r\^{o}le in microscopic physics. However, due to the redefinition of
matter fields, the microscopic physics (\emph{e.g.}, field
theoretic) itself is now described by some modified action
$\tilde{S}_{m}$, which will generally include new interactions
between fundamental particles.

Throughout we have endeavored not to ascribe the label `physical'
to either metric or energy-momentum tensor, although in Palatini
theories one should be aware that in the frame of a laboratory
here on Earth it is $g_{ab}$ and $\hat{T}^{a}{}_{b}$, rather than
$\bar{g}_{ab}$ and $\tilde{T}^{a}{}_{b}$ which behave as we would
expect given our initiation for how things work in General
Relativity. Ultimately, all truly measurable, and hence physical,
quantities should be independent of which names one gives to which
metrics or which frame one works in, and so the names which one
gives to the metrics and energy momentum tensors should only be
seen as a guide to intuition rather than having any deeper
meaning.

\subsection{Coarse-graining the Energy Momentum tensor of Particles}

\label{sec:coarse}

We now consider the coarse-grained form of the effective energy
momentum tensor, $\hat{T}^{a}{}_{b}$ when microscopically matter
is clumped into classical particles surrounded by a vacuum. Given
that we have seen that particle motions in GR and Palatini
theories are the same, we should expect the coarse-grained
effective energy momentum to have the form of the energy momentum
tensor for collisionless dust, as it would in General Relativity.
We show this explicitly below. It should be stressed that the
simple calculation present below is not new in the context of
general relativity \cite{avgreview}, however its consequences for
modified source theories have not been appreciated so far.

Defining $\hat{\mathbf{T}}^{ab} = \sqrt{-g} \hat{T}^{ab} +
\Lambda_{\rm eff} g^{ab}$ we have
\begin{eqnarray}
\nabla_{a} \hat{T}^{ab} = 0 &\Leftrightarrow&  \partial_{i}
\hat{\mathbf{T}}^{ib}  = -\partial_{0} \hat{\mathbf{T}}^{0b} -
\Gamma^{b}_{cd}\hat{\mathbf{T}}^{cd}.
\end{eqnarray}
We now average $\hat{T}^{ab}$ over a region with fixed volume
$\mathcal{V}$ and surface $\Sigma$. $\Sigma$ is chosen to lie in
the vacuum region between the particles such that on $\Sigma$ we
have $\hat{\mathbf{T}}^{a}{}_b = 0$. After some algebra we find
\begin{eqnarray}
\int \mathrm{d}^3 \textbf{x} \hat{\mathbf{T}}^{ij} &=& \frac{1}{2}
\frac{\mathrm{d}^2}{\mathrm{d} t^2}\int \mathrm{d}^3 \textbf{x}\,
x^{i}x^{j} \hat{\mathbf{T}}^{00} \label{Tavg} \\
&&+ \frac{1}{2} \int \mathrm{d}^3 \textbf{x}\, \left [ x^{i}x^{j}
(\Gamma^{0}_{cd} \hat{\mathbf{T}}^{cd})_{,0} + 2x^{(i}
\Gamma^{j)}_{cd} \hat{\mathbf{T}}^{cd}\right]. \nonumber
\end{eqnarray}
For non-relativistic particles, in a weak gravitational field we
have $\Gamma^{a}_{bc} \sim \mathcal{O}(\epsilon)$ and each time
derivative introduces a factor of $\epsilon^{1/2}$ so
\begin{equation}
\int \mathrm{d}^3 x \sqrt{-g} \hat{T}^{i}{}_{j} = -\Lambda_{\rm
eff} \mathcal{V} \delta^{i}{}_{j} + \mathcal{O}(\epsilon) \int
\mathrm{d}^3 x\, \hat{\mathbf{T}}^{00}.
\end{equation}
Indeed using the metric found in Section \ref{Sec:ClassPart} and
assuming that the internal structure of the particles is in
equilibrium, Eq.~(\ref{Tavg}) gives to $\mathcal{O}(\epsilon)$ for
particles with masses $M_{(K)}$ and centers of mass
$x_{(K)}^{i}(t)$:
\begin{equation}
\int \mathrm{d}^3 x \mathbf{T}^{ij} = \sum_{(K)} \dot{x}^{i}_{(K)}
\dot{x}^{j}_{(K)} M_{(K)}.
\end{equation}
Where the sum is over all of the particles inside the volume
$\mathcal{V}$. Similarly, one can show that
\begin{equation}
\int \mathrm{d}^3 x \mathbf{T}^{i0} = \sum_{(K)} \dot{x}^{i}_{(K)}
M_{(K)}.
\end{equation}
This is precisely the same as what one finds in general
relativity, which is unsurprising given that, up to a modified
source, the two theories are equivalent, and we have shown that
the modified source does not affect the motion of classical
particles. It is straightforward to show that the same in true in
a cosmological background provided the relative peculiar
velocities of the particles are non-relativistic and the peculiar
gravitational potential is also small. Thus it is clear that the
coarse-grained energy momentum tensor, $\hat{T}^{a}{}_{b}$, is
that of collisionless dust with a cosmological constant
$\Lambda_{\rm eff}$.  On cosmological scales the former is known
to be well approximated by a pressureless perfect fluid energy
momentum tensor.

\subsection{Changing to the Matter Metric}

\label{sec:physicalmetric}

In this subsection we consider the $f(\mathcal{R})$ gravity
theories using the `natural' metric $\bar{g}_{ab}$ and show that
the results agree with those obtained above using the metric
$g_{ab}$ and with those in the literature. The method used here is
however different from the literature -- we start from the
solutions already obtained in \S~\ref{subsect:1_particle} and make
some coordinate transformation to arrive at our new results. This
analysis is relevant because one may want to find the behavior of
the theory in terms of the original matter metric. A similar
analysis could be carried out for the general $f(\mathcal{R},
\mathcal{R}^{ab}R_{ab})$ theory but that is too complicated and
beyond the scope of this work.

For simplicity and ease of comparison with the results in the
literature let us for now define
\begin{eqnarray} \label{AppenA_Def_AB}
\exp(A) \equiv W\exp(2C), \qquad
\exp(B) \equiv \frac{1}{W}
\end{eqnarray}
and again write the metric of the spacetime inside and outside a
static and spherically symmetric particle as
\begin{eqnarray} \label{AppenA_metric_nonphys}
g_{ab}dx^{a}dx^{b} &=& - e^{A(r)}dt^{2} + e^{B(r)}dr^{2} +
r^{2}d\Omega^{2}
\end{eqnarray}
where $A, B$ are functions of the radial coordinate $r$ due to the
symmetry and are given through Eqs.~(\ref{AppenA_Def_AB},
\ref{Eq:R1_spherical_solution1}).

We already know that the matter metric $\bar{g}_{ab}$ is related
with the metric $g_{ab}$ through $g_{ab} = F\bar{g}_{ab}$ so that
the line element is given as
\begin{eqnarray} \label{AppenA_metric_phys1}
d\bar{s}^{2} &=& \bar{g}_{ab}dx^{a}dx^{b}
\nonumber\\ &=&
\frac{1}{F}\left[- e^{A}dt^{2} + e^{B}dr^{2} +
r^{2}d\Omega^{2}\right].
\end{eqnarray}
On the other hand, if we start from the metric $\bar{g}_{ab}$ at
the very beginning, then the spherical symmetry simply requires
that the metric takes a similar form as
Eq.~(\ref{AppenA_metric_nonphys}), that is
\begin{eqnarray} \label{AppenA_metric_phys2}
d\bar{s}^{2} &=& - e^{\bar{A}(\bar{r})}d\bar{t}^{2} +
e^{\bar{B}(\bar{r})}d\bar{r}^{2} + \bar{r}^{2}d\bar{\Omega}^{2}
\end{eqnarray}
where now the coordinates are all barred as they may be different
from the unbarred ones in Eq.~(\ref{AppenA_metric_nonphys}).

It is not difficult to find out the relations between these two
sets of coordinates, the main observation being that
Eqs.~(\ref{AppenA_metric_phys1}, \ref{AppenA_metric_phys2}) should
be equivalent to each other. Because there is no dependence on the
angular coordinates, we could set, from the comparison between
Eqs.~(\ref{AppenA_metric_phys1}, \ref{AppenA_metric_phys2}),
$d\Omega^{2} = d\bar{\Omega}^{2}$ and thus
\begin{eqnarray} \label{AppenA_Relation_r}
\bar{r}^{2} &=& \frac{r^{2}}{F}.
\end{eqnarray}
Similarly we set $dt = d\bar{t}$. Then we just need to use
\begin{eqnarray} \label{AppenA_Relation_AB1}
\frac{1}{F}e^{A(r)} = e^{\bar{A}(\bar{r})},\qquad
\frac{1}{F}e^{B(r)}dr^{2} = e^{\bar{B}(\bar{r})}d\bar{r}^{2}
\end{eqnarray}
and Eq.~(\ref{AppenA_Relation_r}) to find out the relations
between $A, \bar{A}$ and $B, \bar{B}$ and in this way calculate
\emph{explicitly} the matter metric in
Eq.~(\ref{AppenA_metric_phys2}).

To do this, note that $F$ is a function of $\mathcal{T}$ and thus
of $r$. Equivalently it could also be expressed as a function of
$\bar{r}$. This is because there is a relationship between $r$ and
$\bar{r}$ by Eq.~(\ref{AppenA_Relation_r}), \emph{i.e.}, we could
write $\bar{r}(r)$ or $r(\bar{r})$. So once the form of
$f(\mathcal{R})$ is known and $F(r)$ solved as in
Sec.~\ref{subsect:1_particle}, Eq.~(\ref{AppenA_Relation_r}) can
be used to find out $r(\bar{r})$ and $\bar{r}(r)$, then
$F(\bar{r}) = F[r(\bar{r})]$ could be calculated, at least
numerically.

In what follows we shall use a prime (star) to denote the
derivative with respect to $r$ ($\bar{r}$), \emph{i.e.}, $F' =
\frac{dF(r)}{dr}$ and $F^{\ast} = \frac{dF(\bar{r})}{d\bar{r}}$.
Then Eq.~(\ref{AppenA_Relation_r}) can be written as
$$r =
\sqrt{F(\bar{r})}\bar{r} \Rightarrow dr =
\sqrt{F(\bar{r})}(1+\gamma)d\bar{r}$$ where $\gamma \equiv
\bar{r}F^{\ast}/2F$, and from Eq.~(\ref{AppenA_Relation_AB1}) we
finally obtain
\begin{eqnarray} \label{AppenA_Relation_AB2}
\exp\left[\bar{A}(\bar{r})\right] &=&
\frac{1}{F(\bar{r})}\exp\left[A\left(\sqrt{F(\bar{r})}\bar{r}\right)\right],\nonumber\\
\exp\left[\bar{B}(\bar{r})\right] &=&
\exp\left[B\left(\sqrt{F(\bar{r})}\bar{r}\right)\right](1+\gamma)^{2}.
\end{eqnarray}
There are several points to be noted about these results:
\begin{enumerate}
    \item We manage to calculate the metric
Eq.~(\ref{AppenA_metric_phys2}) without explicitly solving the
complicated modified Einstein equation
Eq.~(\ref{Eq:R1_MFE_Physical}). This method, when generalized
appropriately, should be very useful when one deals with the
$f(\mathcal{R}^{ab}R_{ab})$ gravity theories, in which case, as we
discussed in Sec.~\ref{Sect:PalatiniGravity}, the modified
Einstein equation calculated with the physical metric
$\bar{g}_{ab}$ should be very complicated.
    \item Because $F^{\ast}$ is involved in
Eq.~(\ref{AppenA_Relation_AB2}), $\bar{B}$ and thus the metric
$\bar{g}_{\mu\nu}$ could be discontinuous even though $g_{\mu\nu}$
is continuous. This discontinuity happens when there is a sudden
change of energy density distribution, for example from $\rho =
\rho_{0} = \mathrm{const.}$ inside the particle to $\rho = 0$
outside it. Other cases when there will be singularity can be
found in \cite{pal3}. Furthermore, if the size of a particle with
some fixed mass is tiny, then $\gamma$ could be very large, making
the metric $\bar{g}_{ab}$ deviate from $\eta_{ab}$ significantly.
    \item Outside the particle where $F = \mathrm{const.}$, for
    some choices of $f(\mathcal{R})$, \emph{e.g.} $f(\mathcal{R}) =
    \mathcal{R}^{\alpha}$ with $\alpha \geq 1$, $\mathcal{R} = 0$
    is a vacuum solution to Eq.~(\ref{Trace:Eq}) and so
    outside the particle $F = 1, \gamma = 0$. In this case obviously $\bar{r} =
    r$ and $\bar{A} = A, \bar{B} = B$ outside the particle so that
    the spacetime there is exactly Schwarzschild. In
    other cases $F \neq 1, \gamma \neq 0$, the outside spacetime
    is Schwarzschild-de-Sitter. Consequently the conclusion that
    the spacetime outside a particle is Schwarzschild-de-Sitter
    also holds after changing to the matter metric
    $\bar{g}_{ab}$. Indeed the claim in Sec.~\ref{subsect:Masses} that
    for $f(\mathcal{R}, \mathcal{R}^{ab}R_{ab})$
    gravity theories the active gravitational
    mass of a particle is equal to its inertial mass is correct no
    matter we use $\bar{g}_{ab}$ or $g_{ab}$, which is as
    expected because these two are ultimately \emph{equivalent}.
\end{enumerate}

As a further example, we could use the above method to obtain
explicitly the equations that $\bar{A}(\bar{r})$,
$\bar{B}(\bar{r})$ must satisfy. From the second equation in
Eq.~(\ref{AppenA_Relation_AB2}) we have
\begin{eqnarray}
\bar{B}(\bar{r}) &=& B(r) + 2\log(1+\gamma)\nonumber
\end{eqnarray}
the derivative of which with respect to $\bar{r}$ gives
\begin{eqnarray} \label{AppenA_Eqn_B_bar}
\bar{B}^{\ast} &=& B'\frac{dr}{d\bar{r}} +
\frac{2}{1+\gamma}\gamma^{\ast}\nonumber\\
&=& -e^{B}\frac{d}{dr}e^{-B}\sqrt{F(\bar{r})}(1+\gamma) +
\frac{\frac{F^{\ast}}{F} + \frac{\bar{r}F^{\ast\ast}}{F} -
\bar{r}\left(\frac{F^{\ast}}{F}\right)^{2}}{1+\gamma}\nonumber\\
&=& \frac{1}{1+\gamma}\left[\frac{e^{\bar{B}}}{F}(\kappa\rho + V)
\bar{r} + \frac{1-e^{\bar{B}}}{\bar{r}}\right]\nonumber\\
&& + \frac{1}{1+\gamma}\left[\frac{\bar{r}F^{\ast\ast}}{F} +
2\frac{F^{\ast}}{F} -
\frac{3\bar{r}}{4}\left(\frac{F^{\ast}}{F}\right)^{2}\right]
\end{eqnarray}
where in the second step we have used the relations
$$\gamma =
\bar{r}F^{\ast}/2F, \quad dr = \sqrt{F(\bar{r})}(1+\gamma)d\bar{r}$$
\textrm{and} $$B' = -e^{B}\frac{d}{dr}e^{-B},$$ and in the third step we have
used $e^{B} = \frac{1}{W}$ and
Eq.~(\ref{Eq:R1_spherical_solution1}). Similarly from the first of
Eq.~(\ref{AppenA_Relation_AB2}) we have
\begin{eqnarray}
\bar{A}(\bar{r}) &=& A(r) - \log F\nonumber
\end{eqnarray}
the derivative of which with respect to $\bar{r}$ gives
\begin{eqnarray} \label{AppenA_Eqn_A_bar}
\bar{A}^{\ast} &=& A'\frac{dr}{d\bar{r}} -
\frac{F^{\ast}}{F}\nonumber\\
&=& \frac{1}{1+\gamma}\left[\frac{e^{\bar{B}}-1}{\bar{r}} +
\frac{e^{\bar{B}}}{F}(\kappa p_{I} + \kappa p_{A} -
V)\bar{r}\right]\nonumber\\
&& -
\frac{1}{1+\gamma}\left[\frac{3\bar{r}}{4}\left(\frac{F^{\ast}}{F}\right)^{2}
+ 2\frac{F^{\ast}}{F}\right],
\end{eqnarray}
in which we have used $$A' = e^{-A}\frac{d}{dr}e^{A} =
\frac{W'}{W} + 2C', \qquad e^{B} = \frac{1}{W}$$ and
Eq.~(\ref{Eq:R1_spherical_solution1}). The coupled differential
equations Eqs.~(\ref{AppenA_Eqn_B_bar}, \ref{AppenA_Eqn_A_bar})
govern the $\bar{r}$ dependence of $\bar{A}, \bar{B}$, and by
solving them with appropriate boundary conditions we can get
$\bar{A}, \bar{B}$. This is just another method to solve for
$\bar{A}, \bar{B}$. It is not difficult to verify that these
equations are equivalent to Eqs.~(21, 22) listed in
\cite{DEReview} which are derived using the full modified Einstein
equation with the matter metric $\bar{g}_{ab}$,
Eq.~(\ref{Eq:R1_MFE_Physical}). This again shows the equivalence
between these two methods. A similar set of differential equations
could be obtained for the case of Palatini
$f(\mathcal{R}^{ab}R_{ab})$ as well but that is beyond the scope
of this work.

\subsection{Discussion and Summary}

\label{subsect:SummarySect}

Having directly considered both the motion of classical particles and
the coarse-graining of the energy momentum tensor for matter
consisting of classical particles, we are now ready to discuss how the
coarse-grained averaging in the Palatini modified gravity theories
leads to effects which are distinct from what have been claimed using
the na\"{\i}ve averaging.

On the microscopic level, the matter in our Universe, whether it
is dark or baryonic (radiation will be discussed in the next
section), is made up of small particles. Consequently our previous
analysis of the motion of particles in Palatini modified gravity
theories is directly applicable to this setting. As we have seen,
the effect of the Palatini modification to General Relativity is a
change of the \emph{internal configuration} and \emph{structure}
of the particle, while outside the particle the spacetime is
Schwarzschild-de-Sitter with a cosmological constant
$\Lambda_{\mathrm{eff}}$ determined by the model itself. The
motion of these particles in Palatini theories follows the
geodesics which is precisely the same as in general relativity
with $\Lambda=\Lambda_{\mathrm{eff}}$. There are no new extra
\emph{dynamical} degrees of freedom in the Palatini theories,
which in turn means that there is no new \emph{long-range} forces.
Any new effective force must be \emph{non-dynamical} and act only
at points; so it is entirely local and cannot be felt
inter-particles. The closest analogue to this in particle physics
would be Fermi's original proposal for a theory of the weak force.

We could also see this from the viewpoint of averaging. We showed
above in Section \ref{sec:coarse} that if $\mathcal{T}^{a}{}_{b}$
describes matter clumped into classical particles in some vacuum
then, when properly coarse-grained over scales much larger than
the inter-particle separation, the modified source term,
$\hat{T}^{a}{}_{b}$, has the form of the energy momentum tensor
for collisionless dust plus some effective cosmological constant
$\Lambda_{\rm eff}$. This is entirely as one would expect given
our analysis of particle motions in Palatini theories. For $N$
particles each with physical mass $m_p$ in a volume
$V_{\mathrm{tot}}$, the effective energy density of the dust is
$\rho^{\rm eff}_{\rm matter} = Nm_{p}/V_{\mathrm{tot}}$.

Now we compare our above conclusion with that one would make when
using the na\"{\i}ve averaging by considering some specific
situations (for simplicity we take $f(\mathcal{R})$ theory as an
example, the case of $f(\mathcal{R}^{ab}R_{ab})$ is similar).
\begin{enumerate}
    \item Consider the cosmological setting. In the literature it
    has been extensively claimed that if $f(\mathcal{R})$ is
    chosen so that the deviation from $f(\mathcal{R}) =
    \mathcal{R}$ grows at small values of $\mathcal{R}$, \emph{e.g.},
    $f(\mathcal{R}) = \mathcal{R} - \frac{\alpha}{\mathcal{R}}$,
    then the model could lead to a phase of accelerating expansion of the
    Universe at late times which is different from that predicted
    by general relativity plus a cosmological constant. There is
    nothing wrong with this if we can model the matter distribution as
    \emph{smooth} even at the smallest scales. Since in this
    case both $F(\Phi)$ and $V(\Phi)$ vary as $\Phi$ varies
    with time so that from Eq.~(\ref{EinsteinMod:Eq})
    we could expect the evolution to be different from $\Lambda\mathrm{CDM}$.
    However, as we have emphasized several times, it is more
    realistic to model the matter as made up of small particles on
    microscopic scales and only becoming a fluid on large scales
    after coarse-grained averaging. What happens after this
    averaging? Take $F$ for example, according to
    Eq.~(\ref{Eq:Averaging}) we have
    \begin{eqnarray}
    \langle F\rangle &=& F_{0} + [F(\rho_{c}) -
    F_{0}]\frac{NV_{p}}{V_{\mathrm{tot}}}\nonumber
    \end{eqnarray}
    where $F_{0} \equiv F(\rho_{0}) = F(0)$. As $V_{p} \ll
    V_{\mathrm{tot}}$ it can be shown that $[F(\rho_{c}) -
    F_{0}]\frac{NV_{p}}{V_{\mathrm{tot}}} \ll F_{0}$ because $F$
    does not depend on $\rho$ linearly [$F(\rho_{c}) \sim F_{0}
    \sim \mathcal{O}(1)$!], so essentially we have $\langle F\rangle =
    F_{0}$. Similarly $\langle V\rangle = V_{0}$. Thus according
    to Eq.~(\ref{EinsteinMod:Eq}) the model is
    indistinguishable from $\Lambda\mathrm{CDM}$.
    \item In astrophysical environments such as the Solar System,
    the matter density is so low that $NV_{p} \ll
    V_{\mathrm{tot}}$. Again our analysis for the cosmological
    setting applies here, namely the Palatini theories behave
    indistinguishably from General Relativity plus a cosmological
    constant after coarse-grained averaging. This in particular
    means that the Parametrized Post-Newtonian (PPN) parameters we
    measure should be the same as those in General Relativity.
    Note however that in some exceptionally high-density
    astrophysical systems, such as neutron stars, we have $NV_{p} \sim
    V_{\mathrm{tot}}$ and so the na\"{\i}ve averaging may give a
    reasonably good description; in this case one could expect
    the model predictions of Palatini theories and general relativity to be
    different.
    \item As a last example, consider the matter metric
    $\bar{g}_{ab}$ in Eq.~(\ref{AppenA_metric_phys2}).
    Eq.~(\ref{AppenA_Relation_AB2}) tells us that the metric
    function $\bar{A}$ depends on the \emph{local} value of $F$.
    Now suppose we use the na\"{\i}ve averaging. Since the
    function $\bar{A}$ determines the results of the Rebka-Pound
    experiment, we can do two such experiments, one in the normal
    atmosphere and the other in a vacuum chamber. Obviously in
    these two experiments the values of $F$ are very different,
    \emph{i.e.}, $F(\rho_{\mathrm{atm}}) \neq F_{0}$ where
    $\rho_{\mathrm{atm}}$ is the energy density of the earth
    atmosphere and so one may expect that such an experiment could
    distinguish between Palatini theory and general relativity.
    However, according to our above analysis, after coarse-grained
    averaging the relevant value of $F$ in the atmosphere is
    nothing but $F_{0}$ because $NV_{p} \ll V_{\mathrm{tot}}$.
    This means that the above thinking experiments will not work.
\end{enumerate}

So, in conclusion, at classical level the motion of particles,
cosmology and astrophysics in Palatini modified gravity theories
are indistinguishable from the results of general relativity plus
a cosmological constant. However, it must be emphasized that these
are not equivalent theories. As we mentioned above, the internal
structure of a particle in Palatini theories is generally
different from that in general relativity (though we cannot
measure the differences in masses classically). This point will
become relevant when we consider the structures of atoms and
neutron stars in these two types of gravitational theories (see
\S~\ref{Sect:atom}). Furthermore, in Palatini theories the
evolution of a region of space where the natural microscopic
energy momentum tensor, $\mathcal{T}^{a}{}_{b}$, truly was that of
a continuously-distributed pressureless dust would be different
from that of a region with the same coarse-grained density but
where the matter was microscopically contained in small particles.

 \section{The Electromagnetic Field}

\label{sect:EMfield}

In the previous section we have considered the dynamics of
classical particles in Palatini theories. In this section we
investigate how radiation, especially that due to the
electromagnetic (EM) field (\emph{i.e.}, photons), behaves in
these theories. This is particularly important since it is
necessary to understand the propagation of light in order to
correctly interpret cosmological observations.

We start again from the microscopic theory and will be careful
about the averaging procedure. While in the previous sections we
have modelled the classical particles as occupying tiny portions
of the space and in between of them there is vacuum, the
electromagnetic field permeates in the space diffusively and can
be treated as a continuum. Consequently its averaging procedure is
somehow like the na\"{\i}ve one and slightly different from what
we have met for classical particles.

In the $\bar{g}_{ab}$ frame the energy momentum tensor of the EM
field is:
\begin{eqnarray}
\mathcal{T}^{a}_{b\,\mathrm{EM}} &=& - \bar{g}^{ac}\bar{g}^{de}
F_{cd}F_{eb}+ \frac{1}{4}\delta^{a}{}_{b} \bar{g}^{cf}\bar{g}^{de}
F_{cd}F_{ef}.\nonumber
\end{eqnarray}
where $F_{ab} = 2\partial_{[a}A_{b]}$. We begin by considering the
simple case of $f(\mathcal{R})$ theories. In these theories, with
$\Phi = \mathcal{R}$, Eq.~(\ref{Trace:Eq})  gives:
\begin{equation}
f_{,\Phi}(\Phi) \Phi - 2f(\Phi) = 0
\end{equation}
so $\Phi$ takes its constant vacuum value and does not depend on
$F_{cd}$, and Eq.~(\ref{Metric:Eq}) gives:
$$
\sqrt{-g}g^{ab} = \sqrt{-\bar{g}}\bar{g}^{ab} f_{,\Phi},
$$
and so $\bar{g}^{ab} = f_{,\Phi} g^{ab}$. We also have that
$\mathcal{Q}^2 = f_{,\Phi}^2$. Thus the effective energy momentum
tensor, $\hat{T}^{a}{}_{b}$, that appears in the $g_{ab}$-frame
microscopic Einstein equation, Eq.~(\ref{EinsteinMod:Eq}), is:
\begin{eqnarray}
\hat{T}^{a}_{b\,\mathrm{EM}} &=& - g^{ac}g^{de} F_{cd}F_{eb}
\\ &&
+
\frac{1}{4}\delta^{a}{}_{b} g^{cf}g^{de} F_{cd}F_{ef} - \frac{V(\Phi)}{f_{,\Phi}^2(\Phi)}\delta^{a}{}_{b}.\nonumber
\end{eqnarray}
and $V(\Phi)/f_{,\Phi}^2 = {\rm const}$. Apart from the addition
of a vacuum energy term ($-V(\Phi)/f_{,\Phi}^2(\Phi)
\delta^{a}_{b}$), the effective energy momentum tensor in the
$g_{ab}$ frame has precisely the same form as it does in the
$\bar{g}_{ab}$ frame. It follows that in vacuum:
$$
\nabla_{a} F^{ab} = 0
$$
where $\nabla_{a} g_{bc} = 0$ and we have raised the indices of
$F^{ab}$ using $g_{ab}$. The vacuum energy term generates an
effective cosmological constant, but other than that, the combined
system the Palatini microscopic vacuum Einstein Maxwell equations
in both metric frames is precisely the same as it is in GR. There
is therefore no need to consider further the effect of averaging
in these theories, or the way light propagates, in
$f(\mathcal{R})$ theories since the they will be precisely as they
are in unmodified GR. This comes about because in these theories
the two metrics are conformally related, and the electromagnetic
action is conformally invariant. As such the energy-momentum of
the EM field does not source the extra Palatini degree of freedom
encoded by $\Phi$, and the evolution of the EM field is not
affected by the modification of gravity.

More generally, however, in
$f(\mathcal{R},\mathcal{R}^{ab}R_{ab})$ the two metrics $g_{ab}$
and $\bar{g}_{ab}$ would instead be \emph{disformally} related. In
these theories then the microscopic behaviour of the
electromagnetic field will be altered, and hence its large scale
behaviour may also change. It is most straightforward to see how
the EM field behaves on macroscopic scales if we would in a frame
where gravity is certainly no stronger than it is in GR over
microscopic scales \emph{i.e.} the $g_{ab}$ frame. We therefore
consider the form of the energy momentum tensor,
$\hat{T}^{a}{}_b$, in the $g_{ab}$ frame. Because of the
complicated relationship between $g^{ab}$ and $\bar{g}^{ab}$, as
given by Eqs.~(\ref{Metric:Eq}, \ref{Matrix:Eq}), it is not
possible, in general, to write down a simple expression, in terms
of $F_{ab}$, for $\hat{T}^{a}{}_b$. We therefore consider the
relatively simple, but cosmologically interesting case, where the
EM field is microscopically disordered and hence describes a bath
of radiation.

Choosing an appropriate coordinate system $(t, x, y, z)$ so that
at some point $(t_{0},\mathbf{x}_{0})$, $\bar{g}_{ab} =
\eta_{ab}$, we may write the components of this $\mathcal{T}^{ab}$
explicitly, for example,
\begin{eqnarray}
T^{00} &=& \frac{1}{2}\sum_{i=x, y, z}\left(E^{2}_{i} +
H^{2}_{i}\right)\nonumber\\
T^{01} &=& E_{y}H_{z}-H_{y}E_{z}\nonumber\\
T^{11} &=& \frac{1}{2}\sum_{i=y, z}\left(E^{2}_{i} +
H^{2}_{i}\right) - \frac{1}{2}\left(E^{2}_{x} +
H^{2}_{x}\right)\nonumber\\
T^{12} &=& -(E_{x}E_{y}+H_{x}H_{y}).\nonumber
\end{eqnarray}
These are respectively the energy density, heat flux, pressure and
anisotropic stress terms, in which $E, H$ are the strengths of the
electric and magnetic fields.

Since a totally disordered electromagnetic field necessarily, on
average, has no preferred direction, when microscopic fluctuations
in the field are averaged over we must have:
\begin{eqnarray}
\langle E^{2}_{x}\rangle = \langle E^{2}_{y}\rangle = \langle
E^{2}_{z}\rangle,\ \langle H^{2}_{x}\rangle = \langle
H^{2}_{y}\rangle = \langle H^{2}_{z}\rangle\nonumber
\end{eqnarray}
Also
\begin{eqnarray}
\langle E_{y}H_{z}-H_{y}E_{z}\rangle\ =\ (y\rightarrow x)\ =\
(z\rightarrow x) =\ 0\nonumber
\end{eqnarray}
and
\begin{eqnarray}
\langle E_{x}E_{y}\rangle\ =\ \langle E_{y}E_{z}\rangle\ =\ \langle E_{x}E_{z}\rangle\ =\ 0,\nonumber\\
\langle H_{x}H_{y}\rangle\ =\ \langle H_{y}H_{z}\rangle\ =\
\langle H_{x}H_{z}\rangle\ =\ 0\nonumber
\end{eqnarray}
because of the lack of phase relations amongst the different
components of the field strengths in the disordered EM field. The
above results imply that on macroscopic scales
$\mathcal{T}^{a}{}_{b}$ for a disordered EM field behaves like a
perfect fluid with $\rho=3p$ and vanishing heat flux \&
anisotropic stress. This is precisely how EM field is treated
cosmologically in standard GR. However, in Palatini theories,
$\mathcal{T}^{a}{}_{b}$ only represents the energy momentum tensor
of the $\bar{g}_{ab}$ frame. There gravity can be very strong over
small scales, and hence behaves in a significantly different
fashion to how it behaves in GR. We prefer to consider the
$g_{ab}$ frame energy momentum tensor, $\hat{T}_{ab}$, since in
this frame gravity behaves no differently to how it behaves in GR.

How the averaging actually works in the case of Palatini
$f(\mathcal{R},\mathcal{R}^{ab}R_{ab})$ gravity depends on the
microscopic field equations Eq.~(\ref{EinsteinMod:Eq}), and here
comes the difference between the $f(\mathcal{R})$ and
$f(\mathcal{R}^{ab}R_{ab})$ cases.  For convenience we decompose the
symmetric tensors $R_{ab}$ and $\mathcal{T}_{ab}$ respectively as
\begin{eqnarray}
R_{ab} &=& K_{ab} = \Delta\bar{u}_{a}\bar{u}_{b} + \Xi\bar{\xi}_{ab} +
2\bar{u}_{(a}\Upsilon_{b)} + \Sigma_{ab}, \label{Rdecomp}\\
\mathcal{T}_{ab} &=& \rho\bar{u}_{a}\bar{u}_{b} + p\bar{\xi}_{ab}
+ 2\bar{u}_{(a}q_{b)} + \pi_{ab}, \label{Tdecomp}
\end{eqnarray}
where $\bar{u}_{a}$ is the four velocity of the observer
($\bar{g}^{ab}\bar{u}_{a}\bar{u}_{b}=-1$), $\bar{\xi}_{ab} =
\bar{g}_{ab} + \bar{u}_{a}\bar{u}_{b}$ is the projection tensor to
the hypersurface perpendicular to $\bar{u}$, and $\rho, p, q_{a},
\pi_{ab}$ are the energy density, isotropic stress, heat flux and
anisotropic stress. Note that $\bar{u}^{a}\bar{\xi}_{ab} =
\bar{u}^{a}q_{a} = \bar{u}^{a}\pi_{ab}=0$ and
$\mathcal{R}=-\Delta+3\Xi$. Throughout we use the convention that
$R^{a}{}_{b} = g^{ac}R_{cb}$ but $K^{a}{}_{b} = \bar{g}^{ac}K_{cb}
= \bar{g}^{ac}R_{cb}$.

We note that Eq.~(\ref{Metric:Eq}) gives:
\begin{eqnarray}
R &=& R_{ab}g^{ab} = \frac{K_{ab}}{\mathcal{Q}^2}\left(f_{,\Phi}\bar{g}^{ab}  +2 f_{,\chi} K^{ab}\right), \\
&=& \frac{f_{,\Phi} \Phi + 2f_{,\chi}\chi}{\mathcal{Q}^2} = \frac{3\kappa p-\kappa \rho + 2f}{\mathcal{Q}^2},
\end{eqnarray}
where the last two equalities follow from Eq.~(\ref{Trace:Eq}) and
the definitions: $\Phi = K_{ab}\bar{g}^{ab}$, $\chi =
K_{ab}K^{ab}$. Eq.~(\ref{Matrix:Eq}) then gives (where all indices
are raised w.r.t.~$\bar{g}_{ab}$):
\begin{eqnarray}
\kappa \rho &=& \Delta(f_{,\Phi} - 2f_{,\chi}\Delta) + 2f_{,\chi} \Upsilon_{c}\Upsilon^{c} + \frac{1}{2} f, \label{rho1Eqn} \\
\kappa q_{a} &=& (f_{,\Phi} + 2f_{,\chi}(\Xi - \Delta))\Upsilon_{a} \label{qaEqn} + 2f_{,\chi} \Upsilon^{c}\Sigma_{ca}, \\
\kappa p &=& (f_{,\Phi} + 2f_{,\chi}\Xi)\Xi - \frac{f}{2} \label{p1Eqn} + \frac{2f_{,\chi}}{3}\left(\Sigma_{ab}\Sigma^{ab}
- \Upsilon_{c}\Upsilon^{c}\right), \\
\kappa \pi_{ab} &=& (f_{,\Phi} + 4f_{,\chi}\Xi)\Sigma_{ab} \label{piabEqn} + 2f_{,\chi}(\Sigma_{a}{}^{c}\Sigma_{cb} - \Upsilon_{a}\Upsilon_{b}) \nonumber \\ &&- \frac{2f_{,\chi}}{3}\bar{\xi}_{ab}\left(\Sigma_{cd}\Sigma^{cd} - \Upsilon_{c}\Upsilon^{c}\right)
\end{eqnarray}
We also have:
\begin{eqnarray}
\Phi &=& 3\Xi - \Delta, \\
\chi &=& \Delta^2 + 3\Xi^2 - 2\Upsilon_{c}\Upsilon^{c} + \Sigma^{a}{}_{b}\Sigma^{b}{}_{a},
\end{eqnarray}
and from Eq.~(\ref{Trace:Eq}):
\begin{equation}
f_{,\Phi}\Phi + 2f_{,\chi}\chi - 2f = \kappa(3p-\rho).
\end{equation}
We are concerned primarily with the form of the effective energy
momentum tensor $\hat{T}_{ab}$ which is defined by
Eq.~(\ref{EinsteinMod:Eq}). Defining $U_{a} = \lambda \bar{u}_{a}$
and $\xi_{ab} = g_{ab} + U_{a}U_{b}$ so that $U_{a}U_{b}g^{ab} =
-1$ \emph{i.e.},
$$
\frac{\mathcal{Q}^2}{\lambda^2} = f_{,\Phi} - 2f_{,\chi} \Delta
\equiv \frac{1}{a^2},
$$
we decompose $\hat{T}_{ab}$ thus:
\begin{equation}
\hat{T}_{ab} = \tilde{\rho} U_{a} U_{b} + \tilde{p} \xi_{ab} + 2U_{(a}\tilde{q}_{b)} + \tilde{\pi}_{ab}
\end{equation}
where (raising indices with $g^{ab}$) we have $U^{a}\tilde{q}_{a} = 0$, $U^{a}\tilde{\pi}_{ab} = 0$ and $g^{ab}\tilde{\pi}_{ab} = 0$.

We note that $$\kappa \hat{T}^{a}{}_{b} =
\frac{1}{\mathcal{Q}^2}\left(\mathcal{T}^{a}{}_{b} -
V(\Phi,\chi)\delta^{a}{}_{b}\right)$$ and so:
\begin{eqnarray}
\tilde{\rho} &=& \frac{\rho}{\mathcal{Q}^2}  - \frac{2 a^2 f_{,\chi}}{\mathcal{Q}^2}\Upsilon^{a} q_{a} + \frac{f_{,\Phi}\Phi +2f_{,\chi}\chi - f}{2\kappa \mathcal{Q}^2}, \nonumber \\
\tilde{p} &=& \frac{p}{\mathcal{Q}^2} - \frac{2 a^2 f_{,\chi}}{3\mathcal{Q}^2}\Upsilon^{a}  q_{a} - \frac{(f_{,\Phi}\Phi +2f_{,\chi}\chi - f)}{2\kappa\mathcal{Q}^2}, \nonumber \\
\tilde{q}^{a} &=& \frac{\lambda}{\mathcal{Q}^2}\left[f_{,\Phi}q^{a}  + 2f_{,\chi}\Xi q^{a} + 2f_{,\chi} \Sigma^{a}{}_{b}q^{b}\right] \\ &&- \frac{4f_{,\chi}^2 a^2}{\mathcal{Q}^2} \Upsilon^{a}\Upsilon_{c}q^{c}\nonumber
\end{eqnarray}
A similar expression to the above can be found for
$\tilde{\Sigma}_{ab}$ although for our purposes it shall be not be
needed. We note that if the average values of $q_{a}$ and
$\Sigma_{ab}$ vanish, then so do the average values of
$\tilde{q}_{a}$ and $\tilde{\Sigma}_{ab}$.

The symmetries of $\left \langle \mathcal{T}^{a}{}_{b} \right
\rangle$, \emph{i.e.}, $\left \langle q_{a} \right \rangle = \left
\langle \pi_{ab} \right \rangle = 0$ imply that $\left\langle
\tilde{q}_{a} \right \rangle = \left \langle \tilde{\pi}_{ab}
\right \rangle = 0$, and that:
$$
\left \langle \kappa \tilde{\rho} - 3\kappa \tilde{p} \right \rangle = \left \langle \frac{2(f_{,\Phi}\Phi + 2f_{,\chi}\chi - f)}{\mathcal{Q}^2} \right\rangle,
$$
and by Eq.~(\ref{Trace:Eq}) we have $f_{,\Phi}\Phi +
2f_{,\chi}\chi  = 2f$ so:
\begin{equation}
\left \langle \kappa \tilde{\rho} - 3\kappa \tilde{p} \right \rangle  = \left \langle \frac{2f(\Phi,\chi)}{\mathcal{Q}^2}\right \rangle
\end{equation}
The average form of $\hat{T}^{a}{}_{b}$ would only be equivalent
to radiation with some cosmological constant if its trace (as
given by the above equation) was constant. It is evident from the
above equations, that even if we (incorrectly) replace any
quantity $Q$ with $\langle Q\rangle$ in the above equations,
\emph{e.g.}, $\pi_{ab}\rightarrow\langle\pi_{ab}\rangle$,
$\Upsilon_{a}\Upsilon_{b}\rightarrow\langle\Upsilon_{a}\rangle\langle\Upsilon_{b}\rangle$.)
and use the averaged values $\langle\rho\rangle=3\langle p\rangle$
and $\langle q_{a}\rangle=\langle\pi_{ab}\rangle=0$, then we would
find $\langle\Upsilon_{a}\rangle=\langle\Sigma_{ab}\rangle=0$, and
conclude that all averaged quantities involving $\Upsilon_{a}$ or
$\Sigma_{ab}$ also vanish. If we accept this as true, we still
have by Eqs.~(\ref{rho1Eqn}) and (\ref{p1Eqn}):
\begin{eqnarray}
\left \langle \Delta \right \rangle =  H_{-}(\left \langle \kappa \rho - \frac{1}{2}f \right \rangle), \\
\left \langle \Xi \right \rangle = H_{+}(\frac{1}{3}\left \langle \kappa \rho + \frac{1}{2}f \right \rangle)
\end{eqnarray}
where
$$
H_{\pm}(x) =  \mp\left[\frac{f_{,\Phi}}{4f_{,\chi}} -
\sqrt{\frac{f_{,\Phi}^2}{16f_{,\chi}^2} \pm
\frac{x}{2f_{,\chi}}}\right]
$$
It is then clear that generally when $f_{,\chi} \neq 0$, $\left
\langle \Phi \right \rangle = \left \langle 3\Xi - \Delta \right
\rangle \neq {\rm const}$. Similarly $\left \langle \chi \right
\rangle = \left \langle \Delta \right \rangle^2 + 3\left \langle
\Xi \right \rangle^{2} \neq {\rm const}$, for $\rho \neq 0$. Hence
unless $f_{,\chi} = 0$, we have that $\left \langle f(\Phi,\chi)
\right \rangle $ and then also $\left \langle R \right \rangle$
depend on $\rho$ and hence on $\tilde{\rho}$.  This means that in
general $f(\mathcal{R},\mathcal{R}^{ab}R_{ab})$ theories (in
contrast to the $f(\mathcal{R})$ models)  a radiation dominated
Universe will not obey the expansion law $a\propto
t^{\frac{1}{2}}$, and moreover it will not even behave as a
Universe with radiation and some cosmological constant.  These
deviations will be particularly significant for early time
cosmology if  $f(\mathcal{R}, \mathcal{R}^{ab}R_{ab})$ is chosen
to make the deviation from GR more significant at high densities.
This represents the first complication associated with the
behaviour of the EM field in Palatini theories.

In deriving the above, we assumed incorrectly that because $\left
\langle \Upsilon_{a} \right \rangle = \left \langle \Sigma_{ab}
\right \rangle = 0$, we could take the average values of
quantities such as $\Upsilon_{a}\Upsilon^{a}$ to vanish as well.
Even though $\left \langle q_{a} \right \rangle = 0$, in general
for a disorder electromagnetic field:
$$\langle\kappa^{2}q^{a}q_{a}\rangle \sim \mathcal{O}(E^{2}H^{2})\kappa^{2}$$
and
$$\langle\kappa^{2}\pi^{ab}\pi_{ab}\rangle \sim
\mathcal{O}(E^{4}+H^{4})\kappa^{2}$$ are obviously nonzero. It
then follows from Eqs.~(\ref{rho1Eqn} - \ref{piabEqn}) that
$\kappa\rho, \kappa p, \kappa^{2}q^{a}q_{a}$,
$\kappa^{2}\pi^{ab}\pi_{ab}$ and $\kappa^{3}q^{a}q^{b}\pi_{ab}$
are related to the five (scalar) unknowns $\Delta, \Xi,
\Upsilon^{a}\Upsilon_{a}$, $\Sigma_{ab}\Sigma^{ab}$ and
$\Upsilon^{a}\Upsilon^{b}\Sigma_{ab}$. Also unless $q_{a} =
\pi_{ab} = 0$, we do not generally have $\Upsilon_{a}\Upsilon^{a}
= 0, \Sigma_{ab}\Sigma^{ab}$ or
$\Upsilon_{a}\Upsilon^{b}\Sigma_{ab}$. Indeed since
$\Upsilon_{a}\Upsilon^{a},\,\Sigma_{ab}\Sigma^{ab} \geq 0$, the
average values of these quantities could be non-vanishing even
though the average values of $q_{a}$, $\pi_{ab}$, $\Upsilon_{a}$
and $\Sigma_{ab}$ do vanish.

Now in application to cosmology suppose we have $f \sim
\mathcal{R} \sim \Delta \sim \Xi$ ($\sim$ means of order,
$f\sim\mathcal{R}$ simply means that the correction term is
significant), $F\Delta \sim F\Xi \sim 1$. Then it could be seen
that $\langle\Delta\rangle \sim \langle\Xi\rangle \sim
\mathcal{O}(\kappa\rho)$ and
$\langle\Upsilon_{a}\Upsilon^{a}\rangle \sim
\langle\Sigma^{ab}\Sigma_{ab}\rangle \sim
\mathcal{O}(\kappa^{2}\rho^{2})$, so that
$$\langle\Delta^{2}\rangle \sim \langle\Xi^{2}\rangle
\sim \langle\Upsilon_{a}\Upsilon^{a}\rangle \sim
\langle\Sigma^{ab}\Sigma_{ab}\rangle,$$ and the terms involving
$\Upsilon_{a}\Upsilon^{a}, \Sigma^{ab}\Sigma_{ab}$ in Eqs. (\ref{rho1Eqn} - \ref{piabEqn}) are
\emph{equally important} as other modification terms. So when calculating the cosmology
in a radiation dominated Universe we must take them into account. This is the second complication of the Palatini
$f(\mathcal{R}^{ab}R_{ab})$ model.  We would still, however, find that generally $\left \langle R \right \rangle $ depends on $\rho$, signalling a deviation from the GR behaviour, when $f_{,\chi} \neq 0$.

These two complications make the Palatini
$f(\mathcal{R}^{ab}R_{ab})$ model less trivial than its
$f(\mathcal{R})$ counterpart: for the latter the cosmologies for
both radiation and matter dominated Universes are the same as in
$\Lambda\mathrm{CDM}$, while for the former, as we referred to in
above, the radiation dominated Universe could behave rather
differently from $\Lambda\mathrm{CDM}$, even violating the $a
\propto t^{\frac{1}{2}}$ law of expansion. It will be interesting
to consider such models in more details, but due to its complexity
this is beyond the scope of this paper and will be further
investigated elsewhere.

But how does this difference from Palatini $f(\mathcal{R})$
gravity and from the behaviour of classical particles arise? In the
case of classical particles, there are no interactions between
separated particles other than gravity and this is why the
cosmology is like that of GR with a cosmological constant; the
modification to GR simply alters the internal structure of the
particles. The radiation field, as we said above, could be treated
as a continuum and the new $'$interaction$'$ due to the
modification to GR (which acts at a point!) exists everywhere; so
photons feel the modification everywhere and the corresponding
cosmology is changed: this is essentially the argument of a
na\"{\i}ve averaging. The Palatini $f(\mathcal{R})$ is immune to
this effect because in this model the interaction depends on
$\rho-3p$ which is zero identically. In general $f(\mathcal{R},
\mathcal{R}^{ab}R_{ab})$ models the interaction does affect the
propagation of photons.

\section{Constraints from Atomic Physics}

\label{Sect:atom}

In this \S we provide a particular example of how microscopic
physics tightly constrains the properties of Palatini theories
(see also \cite{Olmo2008}). We consider how the Palatini
modification alters energies of the photons that are emitted when
an electron transitions from one atomic energy level to a lower
one. This analysis is particular to $f(\mathcal{R})$ theories. A
similar calculation could be performed for a general
$f(\mathcal{R}, \mathcal{R}^{ab}R_{ab})$ theory, but this would be
significantly more complicated due to the fact that the light
propagation is also altered in these theories.

Consider the total action for a charged fermion field
$\Psi_{\mathrm{F}}$:
\begin{equation}
S_{\rm tot} = \bar{S}_{\rm grav} + S_{\rm m} \label{action1}
\end{equation}
where
\begin{eqnarray}
\bar{S}_{\rm grav} &=& \int \sqrt{-\bar{g}}\mathrm{d}^4 x \frac{1}{2\kappa} f(\mathcal{R}), \\
S_{\rm m} &=& \int \sqrt{-\bar{g}}\mathrm{d}^4 x \left \lbrace
\bar{\Psi}_{\mathrm{F}} \left(\varrho^{a}\bar{\nabla}_{a} - m_0 -
iq\varrho^{a}A_{a}\right)\Psi_{\mathrm{F}} \right.\nonumber\\
&&\left. - \frac{1}{4} \bar{g}^{ab} \bar{g}^{cd}
F_{ac}F_{bd}\right\rbrace,
\end{eqnarray}
where $A_{a}$ is the electromagnetic field, $q$ is the charge of
the particle, $\varrho^{a}$ are the curved space-time analogue of
the Dirac $\gamma$ matrices and $\varrho^{a} \varrho^{b} +
\varrho^{b} \varrho^{a} = 2\bar{g}^{ab}$.

We will perform our calculations in the Einstein frame,
\emph{i.e.}, where the metric is $g_{ab}$.  We do this because, as
we noted above, in this frame the local curvature of spacetime is
certainly of a similar magnitude to that which one would expect
from general relativity. As such the approximation $g_{ab} \approx
\eta_{ab}$ in the frame of a laboratory experiment will be equally
as valid in Palatini theories as it is in standard general
relativity. The same is not necessarily true for the Jordan frame
where $\bar{g}_{ab}$ is the metric. Working in the Einstein frame
should be viewed merely as a computational convenience and should
not be viewed as attaching any special physical meaning to this
frame. The observable quantities we will extract from our
calculation will be independent of the choice of frame.

We now convert the action, as given by Eq.~(\ref{action1}), to the
Einstein frame. In this frame we have $g_{ab} =
f_{,\Phi}\bar{g}_{ab}$, $f_{,\Phi}\Phi = R$, $\sqrt{-g} =
f^{2}_{,\Phi}\sqrt{-\bar{g}}$, and so
\begin{eqnarray}
\sqrt{-\bar{g}}f(\Phi) &=& \sqrt{-\bar{g}} f_{,\Phi}\Phi +
\sqrt{-\bar{g}}\left[f(\Phi) - f_{,\Phi}\Phi\right]\nonumber \\
&=& \frac{1}{f^{\prime\,2}(\Phi)}\sqrt{-g}\left[R - 2\kappa
V(\Phi)\right],
\end{eqnarray}
where again $\Phi = \mathcal{R}$ and $V(\Phi) =
\left(f_{,\Phi}\Phi - f\right)/2\kappa$. Thus Eq.~(\ref{action1})
becomes:
\begin{equation}
S_{\rm tot} = S_{\rm grav} + S_{\rm m}^{\rm (eff)},
\end{equation}
where
\begin{eqnarray}
S_{\rm grav} = \int \sqrt{-g} \mathrm{d}^4 x\,\frac{R}{2\kappa f^{2}_{,\Phi}}, \\
S_{\rm m}^{\rm (eff)} = S_{m} - \int \sqrt{-g} \mathrm{d}^4 x\,
\frac{V(\Phi)}{f^{\prime\,2}(\Phi)}. \label{actionmattereff}
\end{eqnarray}
The total energy momentum tensor associated with variations in
$S_{\rm m}^{\rm eff}$ with respect to $g_{ab}$ is conserved.

We rewrite $S_{\rm m}$ in the Einstein frame as follows: we define
new fields
$$\varphi = f_{,\Phi}^{-3/8} \Psi_{\mathrm{F}}, \ \
\bar{\varphi} = f_{,\Phi}^{-9/8} \bar{\Psi}_{\mathrm{F}},$$ and
$$\gamma^{a} = f_{,\Phi}^{-1/2}\varrho^{a}$$
so that $$\gamma^{a}\gamma^{b} + \gamma^{b}\gamma^{a} = 2
g^{ab}.$$ We then have:
\begin{eqnarray}
S_{\rm m} = &&\int \sqrt{-g}\mathrm{d}^4 x \left \{ \bar{\varphi}
\left[\gamma^{a}\nabla_{a} - m(\Phi) -
iq\gamma^{a}A_{a}\right]\varphi\right. \nonumber\\ &&\left. -
\frac{1}{4} g^{ab} g^{cd} F_{ac}F_{bd}\right\},
\end{eqnarray}
in which $\nabla_{a}$ satisfies $\nabla_{a}g_{bc}=0$ and $m(\Phi)
= m_{0}f_{,\Phi}^{-1/2}$. Varying this action with respect to
$\varphi$, we arrive at the modified Dirac equation obeyed by
electrons in these theories:
\begin{equation}
\gamma^{a}\partial_{a} \varphi = m(\Phi) \varphi + iq \gamma^{a} A_{a}\varphi, \label{Dseqn}
\end{equation}
Varying the action with respect to $\Phi$ gives us:
\begin{equation}
\frac{2f(\Phi) - f_{,\Phi}\Phi}{f_{,\Phi}^{2}} = \kappa
m(\Phi)\bar{\varphi}\varphi. \label{Phi}
\end{equation}
With the equations written in this form, the effect of the
Palatini modifications is manifest: it is to make the mass of the
electron, $m(\Phi)$, a $\Phi$-dependant quantity (see \emph{e.g.}
\cite{mota1, mota2, mota3} for considerations of other local
density dependent quantities). Since the mass of the electron
depends on $\Phi$, it also depends on the local density of matter,
and hence also on the local electron density. In an atom the peak
local electron density generally decreases as the energy of the
orbit decreases, and so the effective mass of the electron will be
different for different energy orbits. This will lead to
potentially detectable deviations from the standard model of
particle physics. We now quantify these deviations.

We wish to consider the energy eigenstates of an electron orbiting
an a hydrogen nucleus. We therefore take $\varphi \propto \exp(-i
E_{\rm e} t)$ where $E_{\rm e}$ is the electron energy, and for
the hydrogen atom $q A_{0} = \alpha_{\mathrm{EM}}/r =-V(r)$ where
$\alpha_{\mathrm{EM}}$ is the fine structure constant. From
Eq.~(\ref{Dseqn}), the electron obeys:
$$
E_{\rm e} \varphi = -i\tilde{\alpha}_k \partial_{k} \varphi -
\frac{\alpha_{\mathrm{EM}}}{r}\varphi + m(\Phi)\beta \varphi,
$$
where $\tilde{\alpha}_{i}\beta = - \beta\tilde{\alpha}_{i}$,
$\beta^2 = 1$ and $\tilde{\alpha}_{(i}\tilde{\alpha}_{j)} =
\delta_{ij}$. We define:
\begin{equation}
\varphi = \left(\begin{array}{c} F(x) \\ iG(x) \end{array}\right),
\end{equation}
and then:
\begin{eqnarray}
\left[E_{\rm e}-m(\Phi) + \frac{\alpha_{\mathrm{EM}}}{r}\right] F = \mathbf{\sigma}\cdot\nabla G, \label{eqnF}\\
\left[E_{\rm e}+m(\Phi) + \frac{\alpha_{\mathrm{EM}}}{r}\right] G
= -\mathbf{\sigma}\cdot\nabla F, \label{eqnG},
\end{eqnarray}
where $\sigma_{i}$ are the Pauli matrices. $E$ is the total energy
of the electron and includes the contribution from the electron
rest mass, we therefore write $E_{\rm e} = m_{0} + E_{\rm
e}^{\prime}$.

We calculate the modifications to the energies of the photon that
is emitted when an electron changes from one energy level to
another by assuming that the Palatini modification is small enough
so that we may write $f(\Phi) = \Phi[1+\varepsilon(\Phi)]$ where
$\varepsilon(\Phi) \ll 1$.  Not all Palatini theories can be
written in this may, but considering we will find that any changes
in $\varepsilon(\Phi)$ are constrained to be very small, we
believe that is it highly unlikely that any theory, which cannot
be written thus, could be experimentally viable.

Assuming this form for $f(\Phi)$ we have
$$m(\Phi) = m_{0}/\sqrt{f_{,\Phi}} \approx m_{0}[1 -
\frac{1}{2}(\epsilon + \Phi \epsilon_{,\Phi})] \equiv m_{0} +
\delta m(\Phi),$$ where $\delta m(\Phi)/m_{0} \ll 1$.  We solve
Eqs.~(\ref{eqnF}, \ref{eqnG}) perturbatively in the small
parameter $\delta m(\Phi)/m_0$.

To $\mathcal{O}((\delta m/m_0)^{0})$ we have
$$F = \bar{F},\ \ \ G= \bar{G} \quad \textrm{and} \quad E_{\rm e}^{\prime}
= \bar{E}_{\rm e} = \alpha_{\mathrm{EM}}^2 m_{0}
\mathcal{E}(\alpha_{\mathrm{EM}})$$ where having defined $y =
x/a_{0}$, $a_0 = 1/\alpha_{\mathrm{EM}}m_0$:
\begin{eqnarray}
\alpha_{\mathrm{EM}}\left[\mathcal{E} + \frac{1}{\vert y \vert}\right] \bar{F} &=& \mathbf{\sigma}\cdot\nabla_{(y)} \bar{G}, \label{eqnF2}\\
\left[2 + \alpha_{\mathrm{EM}}^2\mathcal{E} +
\frac{\alpha_{\mathrm{EM}}^2}{\vert y \vert}\right] \bar{G} &=&
-\alpha_{\mathrm{EM}}\mathbf{\sigma}\cdot\nabla_{(y)} \bar{F},
\label{eqnG2}
\end{eqnarray}
thus $\bar{G} \sim \mathcal{O}(\alpha_{\mathrm{EM}} \bar{F})$. The
unmodified energy levels, $\bar{E}_{\rm e}$ are just what they
would be in the standard model with an electron mass $m_0$. We do
not repeat a full calculation of those energy levels here. For the
purposes of calculating the large contribution perturbation to the
energy levels we need only work to leading order in
$\alpha_{\mathrm{EM}}$. To the leading order in
$\alpha_{\mathrm{EM}}$, $\bar{F}$ satisfies the Schr\"{o}dinger
equation:
\begin{equation}
-\frac{1}{2}\nabla_{(y)}^2 \bar{F} = \mathcal{E} + \frac{1}{\vert
y \vert} \bar{F}(1+\mathcal{O}(\alpha_{\mathrm{EM}}^2)).
\end{equation}
To this order,  for a state with energy level $n$ and angular
momentum $(lm)$ we have to leading order in
$\alpha_{\mathrm{EM}}$:
$$\mathcal{E} = \mathcal{E}_{n} = -1/2n^2
(1+\mathcal{O}(\alpha_{\mathrm{EM}}^2))$$ and we write
$$\bar{F}_{nlm}  = \bar{R}_{nl}(r)Y_{lm}(\theta, \phi)$$
where $Y_{lm}(\theta,\phi)$ are spherical harmonics, $r$ is the
distance from the nucleus and $\theta$ and $\phi$ are angular
coordinates. We normalize $\bar{F}_{nlm}$ so that:
$$
\int d^3 x \bar{F}^2_{nlm} = 1.
$$
To leading order in both $\alpha_{\mathrm{EM}}$ and $\delta m/m_0$
the electron number density, $n_e$, is given by $n_{e} =
\bar{F}^2$ and, so to this order by Eq.~(\ref{Phi}), where for a
state with energy level $n$ and angular momentum $(lm)$, we have
$\Phi = \bar{\Phi}_{nlm}$ and
\begin{equation}
\bar{\Phi} = \kappa m_{0}\bar{F}^2_{nlm}. \label{Phi2}
\end{equation}
To $\mathcal{O}(\delta m/m_0)$ we write
$$F = \bar{F} + \delta F,\ \ \ G = \bar{G} + \delta G \quad \textrm{and}\quad E_{\rm e}^{\prime} = \bar{E}_{\rm e} + \Delta m.$$
We therefore have:
\begin{eqnarray}
&& -\nabla_{(y)}^2 \delta F\nonumber\\
&=& \sigma \cdot \nabla_{(y)} \frac{\delta m}{m_0}
\frac{\bar{G}}{\alpha_{\mathrm{EM}}} + \left[\frac{\Delta m +
\delta m}{m_{0}}\right] \left(\mathcal{E} + \frac{1}{\vert
y\vert}\right)\bar{F}\nonumber\\ && +
2\alpha_{\mathrm{EM}}^{-2}\left[\frac{\Delta m - \delta
m}{m_0}\right]\bar{F} + \left(\mathcal{E} + \frac{1}{\vert y
\vert}\right) \delta F,
\end{eqnarray}
and so $\delta G \sim \mathcal{O}(\alpha_{\mathrm{EM}} \delta F)$.
To leading order in $\alpha_{\mathrm{EM}}$ then we have $\delta F
= \lambda \bar{F}/\alpha_{\mathrm{EM}}^2$ where:
\begin{equation}
-\nabla_{y}(\vert \bar{F} \vert^2 \nabla_{(y)} \lambda) =
2\left[\frac{\Delta m - \delta m}{m_0}\right]\vert \bar{F} \vert^2,
\end{equation}
integrating this expression gives, for a state with an energy level $n$, and angular momentum $(lm)$:
\begin{equation}
\Delta m_{nlm} = \int \mathrm{d}^3 \mathbf{x}\, \delta m_{nlm}
\vert \bar{F}_{nlm} \vert ^2 = -\frac{m_0}{2} \left\langle
\epsilon + \epsilon_{,\Phi}\bar{\Phi} \right \rangle_{nlm}.
\end{equation}
where $\delta m_{nlm} = \delta m(\bar{\Phi}_{nlm})$ and we have defined:
$$
\left \langle Q \right \rangle_{nlm} = \int \mathrm{d}^3
\mathbf{x} Q(x) n_{e\,(nlm)}(x).
$$

In a local inertial frame we ignore the energy stored in the
gravitational field, however we can see from
Eq.~(\ref{actionmattereff}) that even when $g_{ab} = \eta_{ab}$,
the effective matter action includes a contribution from the
effective potential $V(\Phi)$.  Only when the contribution from
$V(\Phi)$ is included is the energy and momentum conserved with
respect to $g_{ab}$.  Thus (excluding the energy of the nucleus
which we assume to be independent of the electron energy level),
the total conserved energy in this set-up is given by:
$$
E_{\rm tot} = E_{\rm e} + E_{V(\Phi)}+ E_{\gamma},
$$
where $E_{\gamma}$ is the energy stored in the photon field and when $g_{ab} = \eta_{ab}$:
\begin{equation}
E_{V(\Phi)} = \int \mathrm{d}^3 \mathbf{x}\,
\frac{V(\Phi)}{f_{,\Phi}^{2}}.
\end{equation}
Assuming as we have that $f(\Phi) = \Phi(1+\epsilon(\Phi))$, we have:
$$
\frac{V(\Phi)}{f_{,\Phi}^{2}} \approx \frac{\Phi^2
\epsilon_{,\Phi}(\Phi)}{2\kappa}.
$$
Thus taking $\Phi = \bar{\Phi}_{nlm}$ we have:
\begin{equation}
E_{V(\Phi)} = \frac{m_0}{2}\left\langle \epsilon_{,\Phi}\bar{\Phi}^2 \right\rangle_{nlm}.
\end{equation}
where we have used  Eq.~(\ref{Phi2}). Thus:
$$
E_{\rm e} + E_{V(\Phi)} = m_{0} + \bar{E}_{\rm e}(nlm)
-\frac{m_{0}}{2} \left\langle \epsilon \right
\rangle_{nlm}(1+\mathcal{O}(\alpha_{\mathrm{EM}}^2)).
$$
Since $E_{\rm tot}$ is a conserved quantity in this frame, when an
electron changes from a level with quantum numbers $(nl)$ to a
lower energy level with quantum numbers $(n^{\prime} l^{\prime})$,
a photon with energy, $E_{\gamma}$ must be emitted where:
\begin{eqnarray}
E_{\gamma} &&= \bar{E}_{\rm e}({\rm nl}) - \bar{E}_{\rm e}({\rm n^{\prime}l^{\prime}})\\
&&-m_{0} \left[ \left\langle \epsilon \right \rangle_{\rm nl}-
\left\langle \epsilon \right \rangle_{\rm
n^{\prime}l^{\prime}}\right](1+\mathcal{O}(\alpha_{\mathrm{EM}}^2)).\nonumber
\end{eqnarray}
where we have defined:
$$
\left \langle Q \right \rangle_{nl} = \frac{1}{2l+1}\sum_{m = -l}^{l} \left \langle Q \right \rangle_{nlm}.
$$
We define $\bar{E}_{\gamma}({\rm nl, n^{\prime}l^{\prime}})$ be
the energy of the photon to zeroth order in $\epsilon$. We then
have:
\begin{equation}
E_{\gamma} = \left[1 + \Delta_{\rm nl}^{\rm n^{\prime}l^{\prime}}\right]
\bar{E}_{\gamma}({\rm nl, n^{\prime}l^{\prime}}),
\end{equation}
where, using $\bar{E}_{\rm e}({\rm nl})=-\alpha_{\mathrm{EM}}^2
m_0/2n^2$ to leading order in $\alpha_{\mathrm{EM}}$, we have to
leading order in $\alpha_{\mathrm{EM}}$:
\begin{equation}
\Delta_{\rm nl}^{\rm n^{\prime}l^{\prime}} = \frac{\left\langle
\epsilon \right \rangle_{\rm nl}-\left\langle \epsilon \right
\rangle_{{\rm
n^{\prime}l^{\prime}}}}{\alpha_{\mathrm{EM}}^2(n^{-2} -
n^{\prime\,-2})}.
\end{equation}
If $\Delta$ were the same for all transitions then this could be
account for simply by a slight alternation of the electron mass.
To constrain $\Delta$ we consider therefore how the energy of a
photon released due to one transition changes relative to that
emitted in another transition; this is independent of the electron
mass i.e. we consider the ratio of $E_{\gamma}$ for one transition
with that for another. It is also independent of one's choice of
frame. If $t$ is the time in a local inertial frame in the
Einstein frame, and $\bar{t}$ the time in a local inertial frame
(LIF) defined with respect to the Jordan metric, then, since
$g_{ab} = f_{,\Phi} \bar{g}_{ab}$, we have $\mathrm{d} t =
f_{,\Phi}\mathrm{d} \bar{t}$. Thus if a photon has energy
$E_{\gamma}$ in LIF of the Einstein frame metric, $g_{ab}$, it has
energy $E_{\gamma} f_{,\Phi}^{1/2}$ in a LIF of the Jordan frame
metric, $\bar{g}_{ab}$. If the energies of two photons are
measured at the same place, the ratio of those two energies is a
frame independent quantity, as the frame dependant scaling factors
($f_{,\Phi}^{1/2}$) cancel. Thus, even though we have performed
our calculation in the Einstein frame, the quantity which we will
constrain is independent of this frame choice.

We now consider two transitions for where measurements have been
shown to agree to the standard theoretical prediction to a high
accuracy \cite{rydberg}.  Firstly we have the $1S_{1/2}-2S_{1/2}$
transition. For this transition $(nl) = (20)$ and
$(n^{\prime}l^{\prime}) = (10)$. Secondly, we consider the
$2S_{1/2}-8D_{5/2}$ transition.  For this transition $(nl) = (83)$
and $(n^{\prime}l^{\prime}) = (20)$.  Thus:
\begin{eqnarray}
&&\frac{E_{\gamma}(83;20)
\bar{E}_{\gamma}(20;10)}{E_{\gamma}(20;10)
\bar{E}_{\gamma}(83;20)}-1\nonumber\\
&=& - \frac{64}{15\alpha_{EM}^2}\left[\left\langle \epsilon \right
\rangle_{\rm 83}-\left\langle \epsilon \right \rangle_{\rm
20}\right] + \frac{4}{3\alpha_{\mathrm{EM}}^2}\left[\left\langle
\epsilon \right
\rangle_{\rm 20}-\left\langle \epsilon \right \rangle_{\rm 10}\right]\nonumber\\
&=& \frac{28}{5\alpha_{\mathrm{EM}}^2}\left[\left\langle \epsilon
\right \rangle_{\rm 20}- \frac{16}{21}\left\langle \epsilon \right
\rangle_{\rm 83} - \frac{5}{21}\left\langle \epsilon \right
\rangle_{\rm 10}\right].
\end{eqnarray}
The measurements of Ref.~\cite{rydberg} provide the following
constraint: the magnitude of the left hand side of the above
equation to be smaller than $8 \times 10^{-10}$ and so:
\begin{equation}
\left \vert \left\langle \epsilon \right \rangle_{\rm 20}-
\frac{16}{21}\left\langle \epsilon \right \rangle_{\rm 83} -
\frac{5}{21}\left\langle \epsilon \right \rangle_{\rm 10}\right\vert < 8 \times 10^{-16}.\label{bound}
\end{equation}
This represents a very strong constraint on the properties of
Palatini $f(\mathcal{R})$ theories. In particular we can see that,
writing $f(\Phi) \approx \Phi(1+ \epsilon(\Phi))$ when
$\Phi/\kappa$ is of the order of the electron cloud density in
hydrogen, then changes in $\epsilon(\Phi)$ are constrained to be
very small. If we expand $\epsilon$ about some appropriate value
of $\Phi$ and find that to linear order we have
$$\epsilon(\Phi) \approx {\rm const} +
\epsilon_{0}\Phi/ b H_0^2,$$ where $H_{0}^2$ is the value of the
cosmological constant today, Eq.~(\ref{bound}) gives the very
strong constraint: $$\vert \epsilon_{0} \vert \approx \vert
f^{\prime \prime}(\Phi)H_{0}^2/f^{\prime}(\Phi) \vert \lesssim 4
\times 10^{-40}.$$

\section{Discussions and Conclusions}

\label{Sect:summary}

Much of our intuition about how the microscopic behaviour of
gravity affects physics on large scales is based upon Einstein's
general relativity. In this \emph{article} we show that such
intuition \emph{cannot} simply be applied to modified gravity
theories without a detailed analysis of the energy-momentum
microstructure. Indeed, na\"{i}vely averaging over the microscopic
structure will generally lead one to make incorrect predictions,
and inaccurate conclusions as to the validity of the theory. In
particular, the na\"{i}ve averaging procedure is \emph{invalid} in
Palatini theories.

For classical particles, we show that the relative motion of
particles in Palatini theories is indistinguishable from that
predicted by GR plus a cosmological constant. This means that the
cosmology and astrophysics (except in some extreme environments
such as neutron stars) of Palatini $f(\mathcal{R},
\mathcal{R}^{ab}R_{ab})$ models are identical to that of GR plus a
cosmological constant. The above result can also been shown using
our correct averaging procedure, as given in \S \ref{sec:coarse}.
This particularly means that the fine tuning problems associated
with the cosmological constant are not alleviated in Palatini
theories. It should be stressed that although the Palatini
theories predict the same cosmology and astrophysics as GR, they
are completely different theories: not only because they predict
different internal structures of particles, but also because they
behave differently from GR in the presence of electromagnetic
field and in the atomic physics.

When coming to electromagnetic fields in Palatini theories, things
becomes a bit different. In contrast to classical particles, which
are tiny clumps of energy density in between of which there is
vacuum, the EM field permeates in the space and the na\"{\i}ve
averaging actually \emph{works}. However, when performing the
averaging one should also take into account the fact that the
field equations are microscopic and that at microscopic level the
EM field is random and disordered. This make the Palatini
$f(\mathcal{R}^{ab}R_{ab})$ model less trivial than its
$f(\mathcal{R})$ counterpart: for the latter the cosmologies for
both radiation and matter dominated Universes are the same as in
$\Lambda\mathrm{CDM}$, while for the former the radiation dominated Universe could behave rather
differently from $\Lambda\mathrm{CDM}$, even violating the $a
\propto t^{\frac{1}{2}}$ law of expansion.
This difference from Palatini $f(\mathcal{R})$
gravity and from the behaviour of classical particles arises because, in the
case of classical particles, there are no interactions between
separated particles other than gravity. This is why the
cosmology is like that of GR with a cosmological constant. The
modification to GR simply alters the internal structure of the
particles. The radiation field can be treated
as a continuum and the new $'$interaction$'$ due to the
modification to GR (which act at point!) exists everywhere. So
photons feel the modification everywhere and the corresponding
cosmology is changed. The Palatini $f(\mathcal{R})$ is immune to
this effect because in this model the interaction depends on
$\rho-3p$ which is zero identically. In general $f(\mathcal{R},
\mathcal{R}^{ab}R_{ab})$ models the interaction does affect the
propagation of photons. In
summary, the Palatini modifications can affect the propagation of
photons (EM field) and even change the cosmic expansion during
radiation domination. Observational data on Big Bang
Nucleosynthesis could then place some constraints on these models.

Interestingly, although Palatini $f(\mathcal{R})$ theories were
designed to modify gravity on large scales, they actually modify
physics on the smallest scales (\emph{e.g.}, the energy levels of
electrons) while leaving the larger scales practically unaltered.
We show that the observables in atomic physics, such as the energy
levels, can be very sensitive to the Palatini modification to GR,
and indeed experimental data places extremely
stringent constraints on any deviation from GR. In general, before
considering any astrophysical consequences of a modified gravity
theory, it is important then to check that it does not make
unrealistic predictions for atomic physics.

One may wonder why the same averaging problem does not appear in
the metric $f(R)$ gravity models. The answer is that, within the
metric approach, averaging over microscopic scales is generally no
less straightforward than it is in GR; this is because in both
cases all degrees of freedom are dynamical. These dynamics
generally ensure that the field equations, for all degrees of
freedom, are approximately linear for small-scale structures. In
contrast, averaging in Palatini models is not so trivial since the
new degree of freedom is non-dynamical, and so its field equation
remains non-linear even on the smallest scales. This non-linearity
introduces an averaging problem that is specific to Palatini
theories \footnote{$G^{\mu}{}_{\nu}$ depends non-linearly on
$g_{\mu\nu}$, hence averaging over cosmological
 scales is also an issue GR \cite{avgreview}.}. And this explains why the
cosmological behaviour of these theories can be very different
from what has been suggested in the literature \cite{pal1, olmo,
dom}.

It is usually the case that modified gravity theories, other than those considered here, predict extra non-linear terms in the microscopic Einstein equations. It is therefore both important and interesting to check that averaging, both over microscopic and cosmological scale structures, does not significantly alter the macroscopic behaviour of the theory from that which might have otherwise been expected from the microscopic equations. As much as the importance of any back-reaction from averaging remains an open problem in General Relativity, as we have illustrated in this work, it is likely to be an even more significant problem in modified theories of gravity.

\begin{acknowledgments}
The authors thank G. Olmo, E. Flanagan, N. Kaloper and D.
Puetzfeld for discussions and comments. BL acknowledges supports
from the Overseas Research Studentship, Cambridge Overseas Trust
and DAMTP. DFM acknowledges the Humboldt Foundation. DJS
acknowledges STFC funding.
\end{acknowledgments}

\appendix


\begin{thebibliography}
\bibitem{} \expandafter\ifx\csname natexlab\endcsname\relax

\fi \expandafter\ifx\csname bibnamefont\endcsname\relax

\fi \expandafter\ifx\csname bibfnamefont\endcsname\relax

\fi \expandafter\ifx\csname citenamefont\endcsname\relax

\fi \expandafter\ifx\csname url\endcsname\relax

\fi \expandafter\ifx\csname urlprefix\endcsname\relax

\fi \providecommand{\bibinfo}[2]{#2} \providecommand{\eprint}[2][]{\url{#2}}


\bibitem{carroll} S.~M.~Carroll \emph{et al}., Phys.~Rev.~D {\bf 70} 043528 (2004);  T.~P.~Sotiriou and V.~Faraoni,
  arXiv:0805.1726 [gr-qc];  S.~Capozziello,
  Int.\ J.\ Mod.\ Phys.\  D {\bf 11}, 483 (2002); D.~F.~Mota, J.~R.~Kristiansen, T.~Koivisto and N.~E.~Groeneboom, Mon. Not. R. Astron. Soc. 382, 793-800 (2007),
  arXiv:0708.0830 [astro-ph];   J.~D.~Bekenstein,
  Phys.\ Rev.\  D {\bf 70}, 083509 (2004)
  [Erratum-ibid.\  D {\bf 71}, 069901 (2005)].


\bibitem{skordis1}
  C.~Skordis, D.~F.~Mota, P.~G.~Ferreira and C.~Boehm,
  Phys.\ Rev.\ Lett.\  {\bf 96}, 011301 (2006)

\bibitem{skordis2}
  F.~Bourliot, P.~G.~Ferreira, D.~F.~Mota and C.~Skordis,
  Phys.\ Rev.\  D {\bf 75}, 063508 (2007)

\bibitem{li0}
  B.~Li and J.~D.~Barrow,
  Phys.\ Rev.\  D {\bf 75}, 084010 (2007)

\bibitem{li1}
  B.~Li, J.~D.~Barrow and D.~F.~Mota,
  Phys.\ Rev.\  D {\bf 76}, 044027 (2007)


\bibitem{bean1}
  I.~Laszlo and R.~Bean,
  Phys.\ Rev.\  D {\bf 77}, 024048 (2008)

\bibitem{bean2}
  N.~Agarwal and R.~Bean,
  arXiv:0708.3967 [astro-ph].

\bibitem{li2}
  B.~Li, D.~F.~Mota and J.~D.~Barrow,
  Phys.\ Rev.\  D {\bf 77}, 024032 (2008)

\bibitem{halle}
  A.~Halle, H.~Zhao and B.~Li,
  Astrophys.~J.~Suppl., \emph{in press}
  [arXiv:0711.0958 [astro-ph]].

\bibitem{koivisto1}
  T.~Koivisto and D.~F.~Mota,
  Phys.\ Rev.\  D {\bf 75}, 023518 (2007)

\bibitem{hu1}
  Y.~S.~Song, H.~Peiris and W.~Hu,
  Phys.\ Rev.\  D {\bf 76}, 063517 (2007)

\bibitem{koivisto2}
  T.~Koivisto and D.~F.~Mota,
  Phys.\ Lett.\  B {\bf 644}, 104 (2007)

\bibitem{carroll2}
  S.~M.~Carroll, A.~De Felice, V.~Duvvuri, D.~A.~Easson, M.~Trodden and M.~S.~Turner,
  Phys.\ Rev.\  D {\bf 71}, 063513 (2005)


\bibitem{koivisto3}
  T.~Koivisto and D.~F.~Mota,
  arXiv:0801.3676 [astro-ph].

\bibitem{brook}
  A.~W.~Brookfield, C.~van de Bruck, D.~F.~Mota and D.~Tocchini-Valentini,
  Phys.\ Rev.\  D {\bf 73}, 083515 (2006)
  [Erratum-ibid.\  D {\bf 76}, 049901 (2007)]


\bibitem{clifton}
  T.~Clifton, D.~F.~Mota and J.~D.~Barrow,
  Mon.\ Not.\ Roy.\ Astron.\ Soc.\  {\bf 358}, 601 (2005)

\bibitem{ber}
  E.~Bertschinger and P.~Zukin,
  arXiv:0801.2431 [astro-ph].


\bibitem{cog}
  G.~Cognola and S.~Zerbini,
  arXiv:0802.3967 [hep-th].


\bibitem{hall}
  A.~W.~Brookfield, C.~van de Bruck and L.~M.~H.~Hall,
  Phys.\ Rev.\  D {\bf 74}, 064028 (2006)

\bibitem{sot}
  T.~P.~Sotiriou,
  arXiv:0710.4438 [gr-qc].

\bibitem{bo}
  C.~G.~Boehmer, L.~Hollenstein and F.~S.~N.~Lobo,
  Phys.\ Rev.\  D {\bf 76}, 084005 (2007)

\bibitem{pop}
  N.~J.~Poplawski,
  Class.\ Quant.\ Grav.\  {\bf 24}, 3013 (2007)

\bibitem{amen}
  L.~Amendola, D.~Polarski and S.~Tsujikawa,
  Phys.\ Rev.\ Lett.\  {\bf 98}, 131302 (2007)

\bibitem{capo}
  S.~Capozziello, S.~Carloni and A.~Troisi,
  Recent Res.\ Dev.\ Astron.\ Astrophys.\  {\bf 1}, 625 (2003)

\bibitem{steen}
  O.~E.~Bjaelde, A.~W.~Brookfield, C.~van de Bruck, S.~Hannestad, D.~F.~Mota, L.~Schrempp and D.~Tocchini-Valentini,
  JCAP {\bf 0801}, 026 (2008)

\bibitem{erica}
  D.~A.~Easson, R.~Gregory, D.~F.~Mota, G.~Tasinato and I.~Zavala,
  JCAP {\bf 0802}, 010 (2008)

\bibitem{flan} E.~E.~Flanagan,  Phys.~Rev.~Lett.\  {\bf 92}, 071101
(2004) and Class. Quant. Grav.  {\bf 21}, 3817 (2004).


\bibitem{dirk} D.~Puetzfeld and Y.~N.~Obukhov,
  Phys.\ Rev.\  D {\bf 76}, 084025 (2007)

\bibitem{pal1}  S.~Fay \emph{et~al.}, Phys.~Rev.~D {\bf 75}, 063509
(2007); B.~Li \emph{et~al.}, Phys.~Rev.~D \textbf{76}, 024002
(2007); 
  M.~Borunda, B.~Janssen and M.~Bastero-Gil,
  arXiv:0804.4440 [hep-th];
  F.~Bauer and D.~A.~Demir,
  arXiv:0803.2664 [hep-ph];
  G.~J.~Olmo,
  Phys.\ Rev.\  D {\bf 77}, 084021 (2008);  
T.~Koivisto, Phys.~Rev.~D \textbf{73} 083517 (2006); M.~Amarzguioui et
  al.,
Astron.~Astrophys.~{\bf 454}, 707 (2006); 
G.~J.~Olmo,
  Phys.\ Rev.\  D {\bf 72}, 083505 (2005); B. Li and M.~C.~Chu, Phys.~Rev.~D \textbf{74}, 104010
(2006); 

\bibitem{pal2}
  F.~Bauer and D.~A.~Demir,
  arXiv:0803.2664 [hep-ph].


\bibitem{pal3}
  E.~Barausse, T.~P.~Sotiriou and J.~C.~Miller,
  arXiv:0801.4852 [gr-qc].

\bibitem{pal4}
  S.~Lee,
  arXiv:0801.4606 [gr-qc].

\bibitem{li3}
  B.~Li, J.~D.~Barrow and D.~F.~Mota,
  Phys.\ Rev.\  D {\bf 76}, 104047 (2007)


\bibitem{cham1}
  D.~F.~Mota and D.~J.~Shaw,
  Phys.\ Rev.\  D {\bf 75}, 063501 (2007)

\bibitem{cham2}
  D.~F.~Mota and D.~J.~Shaw,
  Phys.\ Rev.\ Lett.\  {\bf 97}, 151102 (2006)


\bibitem[DEReview(2006)]{DEReview} \bibinfo{author}{\bibfnamefont{E.~J.}~%
\bibnamefont{Copeland}}, \bibinfo{author}{\bibfnamefont{M.}~%
\bibnamefont{Sami}} and \bibinfo{author}{\bibfnamefont{S.}~%
\bibnamefont{Tsujikawa}}, \bibinfo{journal}{Int.~J.~Mod.~Phys.~D} \textbf{%
\bibinfo{volume}{15}}, \bibinfo{pages}{1753} (\bibinfo{year}{2006}).

\bibitem[PalaLetter(2008)]{letter} \bibinfo{author}{\bibfnamefont{B.}~%
\bibnamefont{Li}}, \bibinfo{author}{\bibfnamefont{D.~F.}~%
\bibnamefont{Mota}} and \bibinfo{author}{\bibfnamefont{D.~J.}~%
\bibnamefont{Shaw}}, \bibinfo{eprint}{arXiv:0801.0603} (\bibinfo{year}{2008}).

\bibitem[SurfaceIntMethod(1938)]{surfint} \bibinfo{author}{\bibfnamefont{A.}~%
\bibnamefont{Einstein}}, \bibinfo{author}{\bibfnamefont{L.}~%
\bibnamefont{Infled}} and  \bibinfo{author}{\bibfnamefont{B.}~%
\bibnamefont{Hoffmann}},, \bibinfo{journal}{Ann.~Math.} \textbf{%
\bibinfo{volume}{39}}, \bibinfo{pages}{65} (\bibinfo{year}{1938}).

\bibitem[PartMotion(2005)]{flanmotion} \bibinfo{author}{\bibfnamefont{E.}~%
\bibnamefont{Racine}} and \bibinfo{author}{\bibfnamefont{E.~E.}~%
\bibnamefont{Flanagan}}, \bibinfo{journal}{Phys.~Rev.~D} \textbf{%
\bibinfo{volume}{71}}, \bibinfo{pages}{044010} (\bibinfo{year}{2005}).


\bibitem{Olmo2008}
  G.~J.~Olmo,
  arXiv:0802.4038 [gr-qc].

\bibitem[WillReview(2001)]{WillGravReview} \bibinfo{author}{\bibfnamefont{C.}~%
\bibnamefont{Will}}, \bibinfo{journal}{Living Rev. Rel.} \textbf{%
\bibinfo{volume}{4}}, \bibinfo{pages}{4} (\bibinfo{year}{2001}).
\bibitem{rydberg} Th.~Udem \emph{et al.} {\it Phys.~Rev.~Lett.} {\bf 79}, 2646 (1997); B. de Beauvoir \emph{et al.} (1997) {\it Phys~Rev~Lett.}  {\bf
78}, 440 (1997).

\bibitem{nottoPal} A.~Iglesias \emph{et al.}, Phys.~Rev.~D {\bf 76}, 104001
(2007).

\bibitem{olmo} G.~J.~Olmo {\it Phys.~Rev.~Lett.} {\bf 98}, 061101
(2007).

\bibitem{tolman} R.~C.~Tolman {\it Relativity, Thermodynamics and
Cosmology}, Clarendon Press, Oxford, 1934.

\bibitem{dom} A. Dominguez, D. Barraco, Phys. Rev. D {\bf 70}, 043505 (2004)

\bibitem{mota1}
  D.~F.~Mota and J.~D.~Barrow,
  Mon.\ Not.\ Roy.\ Astron.\ Soc.\  {\bf 349}, 291 (2004)


\bibitem{mota2}
  D.~F.~Mota and J.~D.~Barrow,
  Phys.\ Lett.\  B {\bf 581}, 141 (2004)

\bibitem{mota3}
  J.~D.~Barrow and D.~F.~Mota,
  Class.\ Quant.\ Grav.\  {\bf 20}, 2045 (2003)

\bibitem{avgreview} A.~Krasinski~1996, \textit{Inhomogeneous Cosmological Models},
(Cambridge UP, Cambridge); T. Buchert, arXiv:0707.2153; A.A. Coley et al,
 Phys. Rev. Lett. 95 (2005) 151102; S. Rasanen,
 JCAP 0402 (2004) 003,   J.~Behrend, I.~A.~Brown and G.~Robbers,
  JCAP {\bf 0801}, 013 (2008);   H.~Alnes, M.~Amarzguioui and O.~Gron,
  Phys.\ Rev.\  D {\bf 73}, 083519 (2006);   T.~Buchert,
  Gen.\ Rel.\ Grav.\  {\bf 33}, 1381 (2001), Gen.\ Rel.\ Grav.\  {\bf 32}, 105 (2000).

\bibitem{nogo} E.~Barausse \emph{et al.}, {\tt arXiv:gr-qc/0703132}


\end{thebibliography}
\end{document}